\documentclass[journal]{IEEEtran}
\usepackage{amsmath}
\usepackage{amssymb}
\usepackage{amsthm}
\usepackage{bm}
\usepackage{lipsum}
\usepackage{cuted}
\usepackage[section]{placeins}
\usepackage[T1]{fontenc}
\usepackage[dvips]{graphicx}
\usepackage{algorithm}
\usepackage{algorithmic}
\usepackage{slashbox}
\usepackage[labelformat=simple]{subcaption}

\usepackage[numbers,sort,compress]{natbib} 

\usepackage{color}
\usepackage{xcolor}
\usepackage{slashbox}
\usepackage{multirow}
\usepackage{multicol}

\usepackage[normalem]{ulem}

\begin{document}

\title{Spectrum Sharing between High Altitude Platform Network and Terrestrial Network: Modeling and Performance Analysis}
\author{Zhiqing Wei,
        Lin Wang,
        Zhan Gao,
        Huici Wu,
        Ning Zhang,
        Kaifeng Han,
        and Zhiyong Feng
       
\thanks{

}}


\maketitle

\begin{abstract}
Achieving seamless global coverage is one of the ultimate goals of space-air-ground integrated network, as a part of which High Altitude Platform (HAP) network can provide wide-area coverage. However, deploying a large number of HAPs will lead to severe congestion of existing frequency bands. Spectrum sharing improves spectrum utilization. The coverage performance improvement and interference caused by spectrum sharing need to be investigated. To this end, this paper analyzes the performance of spectrum sharing between HAP network and terrestrial network. We firstly generalize the Poisson Point Process (PPP) to curves, surfaces and manifolds to model the distribution of terrestrial Base Stations (BSs) and HAPs. Then, the closed-form expressions for coverage probability of HAP network and terrestrial network are derived based on differential geometry and stochastic geometry. We verify the accuracy of closed-form expressions by Monte Carlo simulation. The results show that HAP network has less interference to terrestrial network. Low height and suitable deployment density can improve the coverage probability and transmission capacity of HAP network.
\end{abstract}


\IEEEpeerreviewmaketitle
\section{Introduction}
A{chieving} seamless global coverage is one of the key goals of the next-generation mobile communication system. To this end, space network, aerial network and terrestrial network will be integrated to provide mobile communication services, yielding the space-air-ground integrated network \cite{6g1,hap1}. As a part of the aerial network, High Altitude Platform (HAP) refers to the aircraft hovering at an altitude of 20-50 km above the ground and maintaining quasi-stationary state. HAP network is between satellite network and terrestrial network, and is higher than Low Altitude Platform (LAP) network. Compared with satellite network, the path loss and delay of HAP network are smaller. Compared with terrestrial network, HAP network has more flexibility in scheduling, larger coverage and is less affected by natural disasters \cite{HAP_advantage}. HAP network will play a critical role in the following scenarios.
 
 \begin{itemize}
 	\item [1)] 
 	\emph {Promoting the deployment of Internet of Things (IoT) devices}: The deployment of HAP is less affected by geographical environment, so that IoT devices supported by HAP network can be deployed in more complex terrains, such as deserts, forests, etc., without the limitations of terrestrial mobile communication systems \cite{HAP_IoT}.       
 	\item [2)]
 	\emph {Achieving seamless coverage}: There exist blind coverage areas in the terrestrial mobile communication systems, especially in the remote areas. Besides, future mobile communication systems will explore high frequency bands. The signals in these frequency bands are easily blocked by obstacles, resulting in holes in the coverage area of terrestrial mobile communication systems. HAP network can be applied as a supplement of mobile communication systems for seamless coverage \cite{HAP_coverage}.
 	\item [3)]
	\emph {Traffic offloading for mobile communication
		systems}: The HAP can flexibly move to a hotspot area to offload the traffic of terrestrial mobile communication systems and ensure the Quality of Service (QoS) of users \cite{HAP_traffic}. 
 \end{itemize}
 
For the sake of global coverage, deploying a large number of HAPs to form a HAP network is indispensable, which will cause severe competition for spectrum resources  with the existing communication systems. Spectrum sharing has been extensively studied in terrestrial and LAP networks to alleviate the problem of spectrum shortage. There are also some efforts on spectrum sharing of HAP network. Sun \emph{et al}. \cite{blockchain} proposed to apply blockchain technology for dynamic spectrum sharing in space-air-ground integrated networks, improving spectrum utilization through the combination of smart contracts and hybrid cloud. Since the functions of HAP and Low Earth Orbit (LEO) satellite are similar, Wang \emph{et al}. \cite{dynamic} proposed a dynamic spectrum sharing and power allocation strategy under the premise of non-ideal spectrum sensing, to maximize the capacity of HAP downlink without affecting the QoS of LEO downlink users. Considering the deployment of HAP at different latitudes, the interference of HAP uplinks to geostationary satellites is studied in \cite{50GHz}, and it is shown that the spectrum sharing between fixed satellite network and HAP network is feasible. 

When HAP network shares spectrum with terrestrial network, the interference of HAPs to the terrestrial network needs to be considered. Alexandre \emph{et al}. \cite{HAP_IMT} studied the coexistence of HAP with point-to-point Fixed Service (FS) in adjacent channels on 2 GHz frequency band and showed that the Interference to Noise Ratio (INR) of HAP to FS is smaller than the given threshold of protection criterion. Mokayef \emph{et al}. \cite{5.8GHz} studied the spectrum sharing at 5.8 GHz, and the results show that the large channel bandwidth and the existence of guard band are beneficial to the coexistence of HAP and FS. Konishi \emph{et al}. \cite{Konishi} proposed a spectrum sharing scheme based on carrier aggregation. Field tests show that in the areas near to Base Station (BS), the interference of HAP will not affect the cellular network. The transmit power of HAP and the separation distance between  terrestrial network and HAP network can be designed to minimize the interference. Jo \emph{et al}. \cite{jo} studied the spectrum sharing between HAP and  BSs, and applied deep reinforcement learning to control the transmit power of HAP to minimize the interruption probability of the HAP downlink to the BSs. Zhou \emph{et al}. \cite{zhou} studied the spectrum sharing between a single HAP and terrestrial network in the 1710-1885 MHz frequency band and obtained the minimum separation distance and power fluxed density mask.

The existing  researches on spectrum sharing between HAP and other networks mainly focus on the perspective of single HAP. There is a lack of the study on the spectrum sharing for HAP network. However, deploying a large amount of HAPs to form the HAP network is indispensable to achieve global coverage. As a powerful tool, stochastic geometry can be applied to model the HAP network and analyze the performance of HAP network \cite{HAP_traffic}. Stochastic geometry is widely applied in cellular network  \cite{BS_PPP1} and Device-to-Device (D2D) network \cite{D2D1}. Besides, a variety of point processes have been applied in LAP network, such as Poisson Point Process (PPP)  \cite{PPP1,PPP3}, Poisson Cluster Process (PCP) \cite{PCP}, Poisson Hole Process (PHP) \cite{PHP}, and Matérn Hard-core Point Process (MHCPP) \cite{MHCPP}. However, since the deployment characteristics of HAPs and LAPs are different, the point processes of LAPs are not directly applicable to model the distribution of HAPs. LAPs operate at a low height and can be approximately regarded as distributed on a 2D plane or in a 3D cuboid. In contrast, HAPs operate at a high height, which causes that the curvature between HAPs and the ground cannot be ignored. Thus, HAPs need to be regarded as distributed on a 2D sphere or in a 3D spherical shell. How to generate the point process on sphere and perform performance analysis is one of the challenges.

This paper applies stochastic geometry to model the distribution of HAP and studies the performance of HAP network and terrestrial network with spectrum sharing. Notably, dense satellite networks have been modeled as PPP on spherical surface in some literature \cite{{PPP on spherical surface 1, PPP on spherical surface 2, PPP on spherical surface 3}}. Different from \cite{{PPP on spherical surface 1, PPP on spherical surface 2, PPP on spherical surface 3}}, we first generalize PPP to regular curves, surfaces, and manifolds. The generalized PPP on curves, surfaces, and manifolds can be converted to PPP on straight lines, planes, and Euclidean spaces borrowing differential geometry. Then we can use traditional PPP for theoretical analysis. However, the converted PPP will become a non-homogeneous PPP. We provide  the simulation steps of scattering points using the sparse  theory. Then, the performance metrics of HAP network with omnidirectional (directional) antenna  and terrestrial network such as coverage probability are analyzed. The main contributions of this paper are as follows.

 \begin{itemize}
	\item [1)] 
	The definition of PPP is generalized to PPP on curves, surfaces and manifolds  to capture the features of HAP networks. For ease of analysis, we use the differential geometry to  convert the  generalization to PPP on straight lines, planes and Euclidean space, respectively. Then, the spectrum sharing between HAP network and terrestrial network is modeled using the generalized PPP theory.  
	\item [2)]
	We analyze the feasibility of spectrum sharing between HAP network and terrestrial network, and derive the coverage probability and transmission capacity of HAP network and terrestrial network. The Monte Carlo simulations verify that HAPs will not cause large interference to terrestrial BSs. Low height and suitable deployment density of HAPs can improve the coverage and transmission capacity of HAP network. Besides, this paper studies the impact of the fluctuations of HAP on the coverage probability of HAP network.
	\item [3)]
	We study the application of directional antenna on HAP network and compare the performance of HAP network with omnidirectional antenna. With directional antenna, the  coverage probability and transmission capacity of HAP network can be improved and the interference to terrestrial network can be reduced.
\end{itemize}

It is noted that parts of this paper have been published in our conference paper \cite{gaozhan}. Compared with the conference version, this paper further generalizes PPP to curves and manifolds, and studies the application of directional antenna to improve the performance of HAP network. Besides, Nakagami-m channel is applied to model air-to-ground channel \cite{whyLoS, nakagami}, which is more realistic than the conference version. We also derive the approximate expression of the coverage probability of HAP network to reduce the computational complexity.

The remainder of this paper is organized as follows. Section II presents the generalization of PPP on curves, surfaces, and manifolds. Section III introduces the system model of spectrum sharing between HAP network and terrestrial network. Sections IV and V apply stochastic geometry and differential geometry to theoretically analyze the performance of HAP network and terrestrial network under omnidirectional and directional antenna, respectively. Section VI verifies the theoretical results by Monte Carlo simulations and Section VII concludes the paper. TABLE \ref{table_1} shows the abbreviations and notations used in this paper.
\begin{table*}[h]
	\caption{\label{sys_para}Abbreviations and Notations}
	\begin{center}
		\begin{tabular}{c l c l}
			\hline
			\hline
			
			{Abbreviation} & {Description}&{Abbreviation} & {Description} \\
			
			\hline
			BS&Base Station&
			CCDF&Complementary Cumulative Distribution Function\\
			D2D&Device-to-Device&
			FS&Fixed Service\\
			HAP&High Altitude Platform&
			INR&Interference to Noise Ratio\\
			IoT&Internet of Things&
			LoS&Line-of-Sight\\
			LAP&Low Altitude Platform&
			LEO&Low Earth Orbit\\
			PCP&Poisson Cluster Process&
			PHP&Poisson Hole Process\\
			PPP&Poisson Point Process&
			PDF&Probability Density Function\\
			QoS&Quality of Service&
			SINR&Signal to Interference plus Noise Ratio\\
			\hline
			\hline
			{Notation} & {Description}&{Notation} & {Description} \\
			
			\hline
			${P_h}$ & Transmit power of HAP&
			${P_g}$ & Transmit power of BS\\
			${P_n}$ & Power of noise  &
			${\alpha_h}$ & Path loss factor of air-to-ground channel\\
			${\alpha_g}$ & Path loss factor of ground-to-ground channel&
			${\lambda_h}$ & Deployment density of HAPs \\
			${\lambda_g}$ & Deployment density of BSs&
			$t$ & HAP deployment angle\\
			$r_b$ & BS deployment radius&
			$\epsilon$ & SINR threshold \\
			$a$ & Earth's radius &
			$h$ & HAP deployment height\\
			$y_0$ & Distance between typical BS user and associated BS&
			$m$ & The shape parameter of Nakagami-m shading\\
			$\sigma^2$ &The scale parameter of Rayleigh fading &
			$\theta_{-3dB}$ & Half-power bandwidth of directional antenna\\
			\hline
			\hline
		\end{tabular}
	\end{center}
	\label{table_1}
\end{table*}

\section{Generalized Poisson Point Process}

In this section, we introduce the definition of generalized PPP on curves, surfaces and manifolds. The generalized PPP can be converted to PPP on straight lines, planes and Euclidean space borrowing differential geometry. The challenge in derivation is to find the mapping relationship between curves, surfaces, manifolds and straight lines, planes, Euclidean spaces, respectively. The converted PPP will generally become non-homogeneous PPP. Thus, it is necessary to use sparse theory to generate the converted PPP, which will be introduced in Section VI. Note that these generalized PPP models also have potential to be applied for the performance analysis of wireless networks distributed on irregular area.

\subsection{PPP on curves}

The definition of one-dimensional PPP is given in \cite[Definition 2.7]{haenggi2012stochastic}. For curves, it is natural to replace bounded interval in the definition of one-dimensional PPP with a part of curve, namely an arc. In the definitions below, we use $\left| . \right|$ to represent the Lebesgue measure that is defined as length of curves and area or hyper volume in high dimensional space \cite{haenggi2012stochastic}.

\textbf{Definition 1}: The  PPP on curve $C$  with density $\lambda(c)$  is a point process satisfying the following two conditions.
\begin{itemize}
	\item [$\bullet$] 
	The arc of curve $C$ is denoted by $c$ and the random variable $N(c)$ follows a Poisson distribution with mean $\int_c {\lambda \left( c \right)\left| dc \right|}$, where $N(c)$ represents the number of points on arc $c$ and $\left| dc \right|$ represents the length of arc element $dc$. The Probability Density Function (PDF) of $N(c)$ is as follows.
		\begin{equation}
			{\rm{Pr}}(N(c) = k) = {{\rm{e}}^{ -  {\int_c {\lambda \left( c \right)\left| dc \right|} }}}\frac{{{{( {\int_c {\lambda \left( c \right)\left| dc \right|} })}^k}}}{{k!}}.
		\end{equation}
	\item [$\bullet$]
	If $c_1$, $c_2$, $\cdots$, $c_m$ are disjoint, $N\left( {c_1} \right)$, $N\left( {c_2} \right)$, $\cdots$, $N\left( {c_m} \right)$ are independent random variables.
\end{itemize}

With \textbf{Definition 1}, we can study the PPP on curves. The basic idea is to convert the PPP on curves to one-dimensional PPP by differential geometry.  A point on the regular curve can be regarded as a point $t$ on real number axis mapped to a point $(x, y, z)$ in  Cartesian coordinates.  Then the curve can be represented by parametric equations with one parameter $t$ \cite{berger2012differential}
\begin{equation}
	\emph{\textbf{r}}(t) = (x(t),y(t),z(t)),
\end{equation}
where $\emph{\textbf{r}}(t)$ represents the vector from the origin to the point on the curve with parameter $t$. The functions $x(t)$, $y(t)$ and $z(t)$ jointly determine the corresponding relationship between the points on real number axis and the points in three-dimensional space. Assuming that there is an arc element $dc$ on the curve, the mean of the number of points on $dc$ is derived as \eqref{point of curves} according to the differential geometry on curves.
\begin{equation}
	dN = \lambda(c) dc = \lambda(t) \left\| {\emph{\textbf{r}}'(t)} \right\|_2dt,
	\label{point of curves}
\end{equation}
where  $\emph{\textbf{r}}'(t)$ is the derivative of $\emph{\textbf{r}}(t)$ and $\left\| {\emph{\textbf{r}}'(t)} \right\|_2$ represents the ${\ell ^2}$-norm (i.e. modulus) of $\emph{\textbf{r}}'(t)$. Then the PPP on curves can be transformed into the one-dimensional PPP, and its equivalent density is
\begin{equation}
	{\lambda ^e(t)} = \frac{{dN}}{{dt}} = \lambda(t) \left\| {\emph{\textbf{r}}'(t)} \right\|_2.
\end{equation}


\subsection{PPP on surfaces}
\begin{figure}[!t]
	\centering
	\includegraphics[scale=0.2]{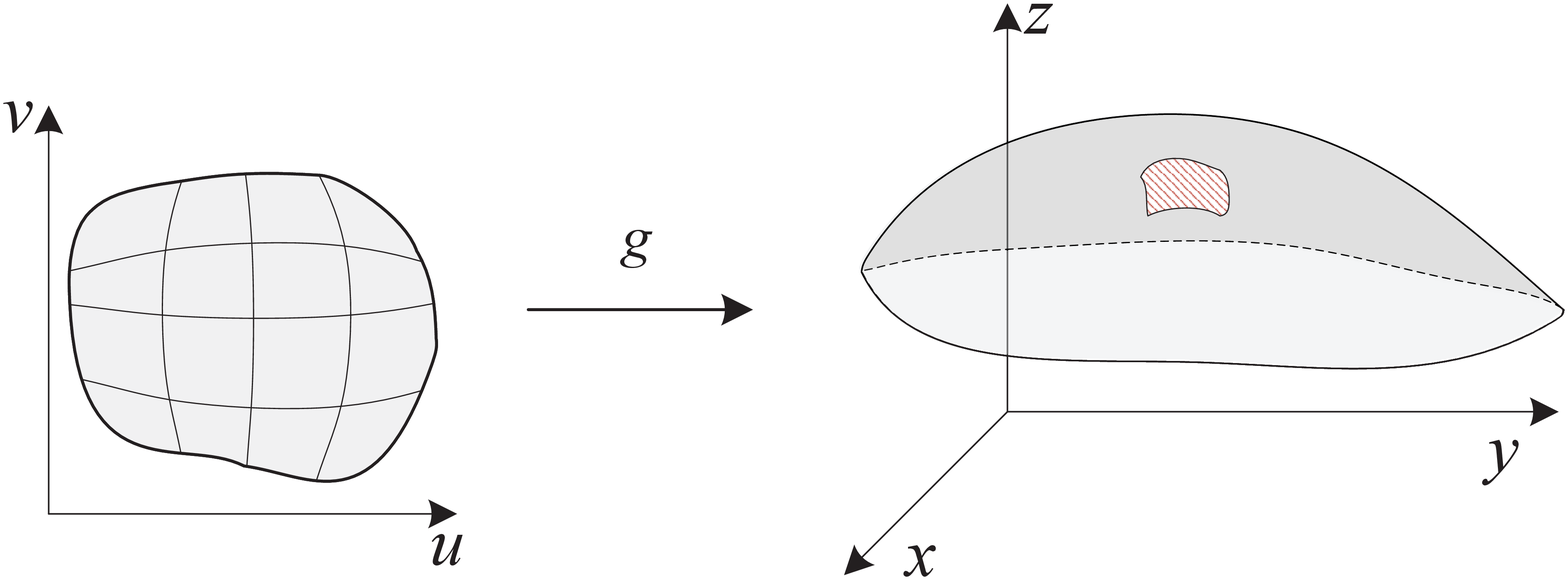}
	\caption{\centering {Mapping \textit{g}$:\left( {u,v} \right) \mapsto \left( {x,y,z} \right)$.}}
	\label{surface}
\end{figure}

The definition of PPP in Euclidean space is given in \cite[Definition 2.10]{haenggi2012stochastic}. For a surface, it is natural to use a part of the surface, namely an area element on surface as a compact set, as shown in Fig. \ref{surface}. We have  introduced the definition of PPP on surface in our previous conference paper \cite{gaozhan}.
%

\textbf{Definition 2}: The PPP on surface $S$ with density $\lambda(s)$  is a point process satisfying the following two conditions.
\begin{itemize}
	\item [$\bullet$] 
	A part of surface $S$ is denoted by $s$ and the random variable $N(s)$ follows a Poisson distribution with mean $\int_s {\lambda \left( s \right)\left| ds \right|}$, where $N(s)$ represents the number of points on $s$ and $\left| ds \right|$ represents the area  of $ds$.  The PDF of $N(s)$ is as follows.
	\begin{equation}
		{\rm{Pr}}(N(s) = k) = {{\rm{e}}^{ -  {\int_s {\lambda \left( s \right)\left| ds \right|} }}}\frac{{{{( {\int_s {\lambda \left( s \right)\left| ds \right|} })}^k}}}{{k!}}.
	\end{equation}
	\item [$\bullet$]
	If $s_1$, $s_2$, $\cdots$ ,$s_m$ are disjoint,  $N\left( {s_1} \right)$, $N\left( {s_2} \right)$, $\cdots$, $N\left( {s_m} \right)$ are independent random variables.
\end{itemize}

Then, we can convert the PPP on surfaces to the PPP on planes for tractability. A point on the regular surface can be regarded as a point $(u, v)$ on two-dimensional plane mapped to a point $(x, y, z)$ in the Cartesian coordinates by mapping \textit{g} as shown in Fig. \ref{surface}.  Then the surface can be represented by parametric equations with two parameters $u$ and $v$ \cite{berger2012differential}
\begin{equation}
	\emph{\textbf{r}}(u,v) = (x(u,v),y(u,v),z(u,v)),
\end{equation}
where $\emph{\textbf{r}}(u,v)$  represents the vector from the origin to the point on the surface with the parameters $(u,v)$. The functions $x(u,v)$, $y(u,v)$ and $z(u,v)$ jointly determine the corresponding relationship between the points on the two-dimensional plane and the points in the three-dimensional space. Supposing that there is an infinitesimal area element $ds$  on the surface, the mean of the number of points on $ds$  is derived as \eqref{point of surfaces} according to the differential geometry on surfaces.
\begin{equation}
	dN = \lambda(s) ds  = \lambda(u,v) \sqrt {EG - {F^2}} dudv,
	\label{point of surfaces}
\end{equation}
where  $E = \left\langle {{\emph{\textbf{r}}'_u}(u,v),{\emph{\textbf{r}}'_u}(u,v)} \right\rangle$, $F = \left\langle {{\emph{\textbf{r}}'_u}(u,v),{\emph{\textbf{r}}'_v}(u,v)} \right\rangle$, $G = \left\langle {{\emph{\textbf{r}}'_v}(u,v),{\emph{\textbf{r}}'_v}(u,v)} \right\rangle$.
Note that ${\emph{\textbf{r}}'_u}(u,v)$ and ${\emph{\textbf{r}}'_v}(u,v)$ represent the derivative function of ${\emph{\textbf{r}}}(u,v)$ with respect to $u$  and $v$,  respectively. And $\left\langle , \right\rangle$ represents the inner product of two vectors.
Then, we can regard the points on surfaces as the points on $u-v$ plane, where the coordinates of points are determined by the parameters $u$ and $v$. In this way, the equivalent density of PPP on $u-v$ plane is
\begin{equation}
	{\lambda ^e(u,v)} = \frac{{dN}}{{dudv}} = \lambda(u,v) \sqrt {EG - {F^2}}.
	\label{equivalent_density}
\end{equation}
Thus, the PPP on planes can be applied to analyze the PPP on surfaces.

\subsection{PPP in manifolds}
A regular surface is actually a simple two-dimensional manifold \cite{do1992riemannian}. Therefore, it is natural to generalize PPP to manifolds.

\textbf{Definition 3}: The $n$-dimensional manifold ${M^n} \subset \mathbb{R}^{n+1}$  is a mapping $f:\mathbb{R}^n \to {M^n}$. The PPP in manifold ${M^n}$  with density $\lambda(v)$ is a point process satisfying the following two conditions.
\begin{itemize}
	\item [$\bullet$] 
	A part of manifold ${M^n}$ or a submanifold is denoted by $v$ and the random variable $N(v)$ follows a Poisson distribution with mean $\int_v {\lambda \left( v \right)\left| dv \right|}$, where $N(v)$ represents the number of points on $v$ and $\left| dv \right|$ represents the hyper volume  of manifolds element $dv$. The PDF of $N(v)$ is as follows.
	\begin{equation}
		{\rm{Pr}}(N(v) = k) = {e^{ - \int_v {\lambda \left( v \right)\left| {dv} \right|} }}\frac{{{{\left( {\int_v {\lambda \left( v \right)\left| {dv} \right|} } \right)}^k}}}{{k!}}.
	\end{equation}
	\item [$\bullet$]
	If $v_1$, $v_2$, $\cdots$, $v_m$ are disjoint, $N( {v_1} )$, $N( {v_2} )$, $\cdots$, $N( {v_m} )$ are independent.
\end{itemize}

Thus, the PPP in manifold is converted to the PPP in Euclidean space $\mathbb{R}^n$. Assuming that the mapping from $\mathbb{R}^{n}$  to $M^n$  is $f\left( {{x_1},{x_2}, \cdots ,{x_n}} \right)$, the mean of the number of points on  manifold element $dv$ is derived as \eqref{point of manifolds} according to the differential geometry on manifolds.
\begin{equation}
	dN = \lambda(v) dv = \lambda(x_1, x_2, \cdots, x_n) \sqrt {\det \left( {{Q}} \right)} d{x_1}d{x_2} \cdots d{x_n},
	\label{point of manifolds}
\end{equation}
where  $\det \left( \cdot \right)$ is the  determinant of a matrix. The element at $i$th row and $j$th column in matrix $Q$ is ${q_{ij}} = \langle {\frac{{\partial f}}{{\partial {x_i}}},\frac{{\partial f}}{{\partial {x_j}}}}\rangle$. In this way, the equivalent density of PPP in Euclidean space $\mathbb{R}^n$ is
\begin{align}
	{\lambda ^e(x_1, x_2, \cdots, x_n)} &= \frac{{dN}}{{d{x_1}d{x_2} \cdots d{x_n}}} \notag \\
	&= \lambda(x_1, x_2, \cdots, x_n) \sqrt {\det \left( {{Q}} \right)}.
\end{align}
Thus, the PPP in Euclidean space $\mathbb{R}^n$ can be applied to analyze the PPP in manifolds.

With the proposed models in this section, the PPP on curves can be converted to the PPP on straight lines, which can be used to model vehicles or road side units on the curved road. The PPP on surfaces can be converted to the PPP on planes for modeling HAPs or satellites distributed at the same altitude. The PPP in manifolds can be converted to the PPP in regular cuboids in three-dimensional space, which can be used to model satellites with different heights distributed in a thick spherical shell. In the following section, we use the generalized PPP  to model HAPs distributed on a spherical surface.

\section{System Model}
\subsection{Network model}
The spectrum sharing between HAP network and terrestrial network is studied in this paper. The BSs of terrestrial network following homogeneous PPP ${\Phi _g}$  with density ${\lambda _g}$  are distributed on earth \cite{BS_PPP1}. HAP network is also a homogeneous PPP ${\Phi _h}$  with density  ${\lambda _h}$ distributed on a sphere with the same center as the earth. The radius $b$ of this sphere is the sum of earth radius $a$ and HAP distribution height $h$. In this paper, we consider the spectrum sharing between the downlink of HAP network and the downlink of  terrestrial network. Without loss of generality, the typical users of HAP and BS are placed in the center of coverage area, whose coordinate is $\left( {0,0,a} \right)$ (Unless otherwise specified, coordinates are Cartesian coordinates in this paper). The corresponding HAP and BS are called typical HAP and typical BS, denoted by $o_h$ and $o_b$, respectively.
\begin{figure}[!t]
	\centering
	\includegraphics[scale=0.5]{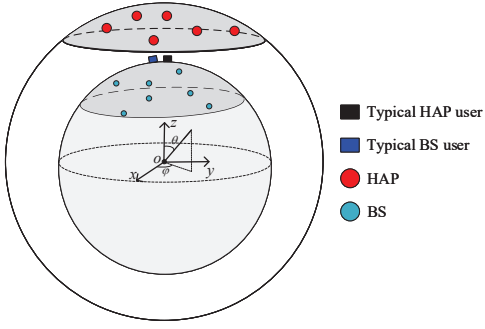}
	\caption{\centering { Distribution model of BSs and HAPs.}}
	\label{Figure_2}
\end{figure}

Because the distribution height of HAP is high, the curvature of earth cannot be ignored. Therefore, the HAPs are assumed to be distributed on a spherical surface. A spherical coordinate system is established with the center being the center of the earth, as shown in Fig. \ref{Figure_2}. $\theta$  and $\phi$  are polar angle and azimuth angle, respectively. In this paper, the Line-of-Sight (LoS) links between HAPs and typical user are considered because the height of HAPs causes large LoS probability \cite{whyLoS}. Thus, the range of parameters is $0 \le \theta  \le t$, $0 \le \varphi  \le 2\pi $. The maximum polar angle $t$ that maintains the LoS between the HAP and typical user is
\begin{equation}
	t = {\cos ^{ - 1}}\left( {\frac{a}{a+h}} \right).
	\label{expression_t} 
\end{equation}
The coordinate of HAP is
\begin{equation}
	\emph{\textbf{r}}\left( {\theta ,\varphi } \right){\rm{ = }}\left( {b\sin \theta \cos \varphi ,b\sin \theta \sin \varphi ,b\cos \theta } \right).
	\label{map}
\end{equation}
Then we can get 
\begin{equation}
	{\emph{\textbf{r}}'_\theta } = \left( {b\cos \theta \cos \varphi ,b\cos \theta \sin \varphi , - b\sin \theta } \right),
\end{equation}
\begin{equation}
	{\emph{\textbf{r}}'_\varphi } = \left( {- b\sin \theta \sin \varphi ,b\sin \theta \cos \varphi , 0 } \right).
\end{equation}

According to \eqref{point of surfaces} \eqref{equivalent_density}, the equivalent density of HAPs on $\theta {\rm{ - }}\varphi $ plane is
\begin{equation}
	\lambda _h^*\left( \theta  \right) = \lambda_h{b^2}\sin \theta.
	\label{the equivalent density}
\end{equation}

Since typical BS user is only affected by nearby BSs, the BSs are approximately distributed in a circular area with a sufficiently large radius $r_b$ centered on the typical user, and the position of BS is expressed in polar coordinates. 

\subsection{Channel model}
\subsubsection{Air-to-ground channel model}
The high deployment of HAP results in a large LoS probability \cite{whyLoS}. Because Nakagami-m channel model captures a wide range of fading scenarios \cite{whynakagami}, this paper applies independent Nakagami-m fading, which has been adopted in \cite{whyLoS, nakagami}, to model  air-to-ground channel  of HAP network. The PDF and Complementary Cumulative Distribution Function (CCDF) of the power gain $h$ are \eqref{nakagami_PDF} and \eqref{nakagami_CCDF} \cite{whyLoS, nakagami}.
\begin{equation}
	p\left( h \right) = \frac{{{m^m}{h^{m - 1}}}}{{\Gamma \left( m \right)}}\exp \left( { - mh} \right),
	\label{nakagami_PDF}
\end{equation}
\begin{equation}
	{\rm{Pr}}\left\{ {h > x} \right\} = \frac{1}{{\Gamma \left( m \right)}}\int_{mx}^{ + \infty } {{y^{m - 1}}{e^{ - y}}} dy \buildrel \Delta \over = \frac{{\Gamma \left( {m,mx} \right)}}{{\Gamma \left( m \right)}}
	\label{nakagami_CCDF},
\end{equation}
where $m$ is the shape parameter of Nakagami-m fading. $\Gamma \left( {m,mx} \right) = \int_{mx}^{ + \infty } {{y^{m - 1}}{e^{ - y}}dy}$ represents the upper incomplete gamma function. $\Gamma \left( m \right)$ represents the gamma function. When $m$ is an integer, $\Gamma \left( {m,mx} \right)$ satisfies
\begin{equation}
	\Gamma \left( {m,mx} \right){\rm{ = }}\left( {m - 1} \right)!\exp \left( { - mx} \right)\sum\limits_{k = 0}^{m - 1} {\frac{{{{\left( {mx} \right)}^k}}}{{k!}}} 
	\label{upper incomplete gamma}.
\end{equation}

\subsubsection{Ground-to-ground channel model}
There are multiple obstacles on the ground, resulting in multiple reflection paths, which can be modeled by independent Rayleigh channel model that is widely applied to model the ground-to-ground channel. The PDF and CCDF of the channel power gain $g$ are shown in  \eqref{rayleigh_PDF} and \eqref{rayleigh_CCDF}, respectively \cite{gaozhan}. 
\begin{equation}
	p\left( g \right) = \frac{1}{{2{\sigma ^2}}}\exp \left( { - \frac{g}{{2{\sigma ^2}}}} \right) 
	\label{rayleigh_PDF},
\end{equation}
\begin{equation}
	{\rm{Pr}}\left\{ {g > x} \right\} = \exp \left( { - \frac{x}{{2{\sigma ^2}}}} \right)
	\label{rayleigh_CCDF},
\end{equation}
where $\sigma$  is the scale parameter of Rayleigh fading.

\section{HAP Network with Omnidirectional Antenna}
In this section, the feasibility of spectrum sharing between HAP network with omnidirectional antenna and terrestrial network is analyzed. The coverage probabilities of HAP network and terrestrial network are derived respectively. And approximate expression for the coverage probability of HAP network is obtained to reduce the computational complexity. The fluctuation of HAPs' height is taken into account when deriving the coverage probability of HAP network.
\subsection{Coverage probability of HAP network}
Coverage probability is expressed in terms of the successful communication probability, which is the probability that the Signal to Interference plus Noise Ratio (SINR) is greater than a given threshold. As HAP network shares the spectrum of terrestrial network, a typical HAP user suffers from the interference of other HAPs and BSs, and its SINR is
\begin{equation}
	\varUpsilon_{h1} = \frac{{{P_h}x_0^{ - {\alpha _h}}{h_0}}}{{ P_n + \sum\limits_{{x_i} \in \Phi_h \setminus \left\{ {{o_h}} \right\}} {{P_h}x_i^{ - {\alpha _h}}{h_i} + \sum\limits_{{y_j} \in {\Phi _g}} {{P_g}y_j^{ - {\alpha _g}}{g_j}} } }},
	\label{sinr_h1}
\end{equation}
where $P_h$ and $P_g$ are the transmit power of HAP and BS, respectively. $x_0$ is the distance from typical HAP user to its associated HAP. $y_j$ and $x_i$ are the distance from $j$th BS and $i$th HAP to the typical HAP user, respectively. $\alpha_h$ and $\alpha_g$ are large-scale path loss coefficients of air-to-ground channel and ground-to-ground channel, respectively. $P_n$ is the power of noise. When SINR is greater than a given threshold $\varepsilon$, the typical HAP user can successfully communicate with the associated HAP. Then, we derive the exact and approximate expressions of  the coverage probability of HAP network as follows. 
 
\textbf{Theorem 1:}  The coverage probability of HAP network when $m$ is an integer is \eqref{coverage_exact_h1}, where ${I_{h1}} = \sum\nolimits_{{\Phi _h}\backslash \left\{ {{o_h}} \right\}} {x_i^{ - {\alpha _h}}{h_i}} $ and ${I_{g1}} = \sum\nolimits_{{\Phi _g}} {y_j^{ - {\alpha _g}}{g_j}} {P_g}/{P_h}$ denote the aggregated interference from other HAPs and BSs to typical HAP user, respectively. ${L_{{I_{h1}}}}\left( s \right)$ (${L_{{I_{g1}}}}\left( s \right)$)  represents the Laplace transform of ${I_{h1}}$ (${I_{g1}}$), as shown in \eqref{Lh} and \eqref{Lg}. And $L_{{I_{h1}}}^{(k)}(s) = \frac{{{\partial ^k}}}{{\partial {s^k}}}{L_{{I_{h1}}}}(s)$ ($L_{{I_{g1}}}^{(k)}(s) = \frac{{{\partial ^k}}}{{\partial {s^k}}}{L_{{I_{g1}}}}(s)$) is the $k$-order derivative function of ${L_{{I_{h1}}}}\left( s \right)$ (${L_{{I_{g1}}}}\left( s \right)$). 
In \eqref{coverage_exact_h1}, the Laplace transform value can be obtained at $s=m\varepsilon x_0^{{\alpha _h}}$. Hence, in the following analysis, we only give the expression of the Laplace transform.
 \begin{figure*}
 	\begin{align}
 		{{\rm{Pr}} _h}\left\{ {{\Upsilon _{h1}} > \varepsilon } \right\} = \exp ( - m\varepsilon x_0^{{\alpha _h}}\frac{{{P_n}}}{{{P_h}}})
 		{\left. {\left\{ {\sum\limits_{k = 0}^{m - 1} {\frac{{{s^k}}}{{k!}}\sum\limits_{l = 0}^k {C_k^l{{\left( {\frac{{{P_n}}}{{{P_h}}}} \right)}^l}\sum\limits_{z = 0}^{k - l} {{{( - 1)}^{k - l}}C_{k - l}^zL_{{I_{h1}}}^{(z)}(s)L_{{I_{g1}}}^{k - l - z}(s)} } } } \right\}} \right|_{s = m\varepsilon x_0^{{\alpha _h}}}}.
 		\label{coverage_exact_h1}
 	\end{align}
	 \begin{equation}
	 	\begin{array}{l}
	 		{L_{{I_{h1}}}}\left( s \right) = \exp \left( { - {\lambda _h}\int_0^{2\pi } {\int_0^t {\left[ {1 - \frac{{{m^m}}}{{{{\left( {m + s{{(\sqrt {{a^2} + {b^2} - 2ab\cos \theta } )}^{ - {\alpha _h}}}} \right)}^m}}}} \right]{b^2}\sin \theta d\theta d\varphi } } } \right).	
 		\end{array}
	 	\label{Lh}
	 \end{equation}
	 \begin{align}
	 	{L_{{I_{g1}}}}\left( s \right) = \exp \left( { - {\lambda _g}\int_0^{2\pi } {\int_0^{{r_b}} {\left[ {1 - \frac{1}{{1 + 2{\sigma ^2}s\frac{{{P_g}}}{{{P_h}}}{y^{ - {\alpha _g}}}}}} \right]ydyd\varphi } } } \right).
	 	\label{Lg}
	 \end{align}
	 \begin{align}
	 	f\left( s \right) = \sum\limits_{k = 0}^{m - 1} {\frac{{{s^k}}}{{k!}}} \sum\limits_{l = 0}^k {C_k^l{{\left( {\frac{N}{{{P_h}}}} \right)}^l}\sum\limits_{z = 0}^{k - l} {C_{k - l}^z{{\left( { - 1} \right)}^{k - l}}\left[ {\frac{{{\partial ^z}}}{{\partial {s^z}}}{L_{{I_{h1}}}}\left( s \right)} \right]\left[ {\frac{{{\partial ^{k - l - z}}}}{{\partial {s^{k - l - z}}}}{L_{{I_{g1}}}}\left( s \right)} \right]} }.
	 	\label{fs_expand}
	 \end{align}
 {\noindent} \rule[-10pt]{18cm}{0.05em}
 \end{figure*}
\begin{proof}
	According to  \eqref{sinr_h1}, the coverage probability of HAP network is expressed as
	\begin{align}
		&{{\rm{Pr}}_h}\{ {\varUpsilon_{h1} > \varepsilon } \}\notag\\
		&= {\rm{Pr}}\left \{ {\frac{{{P_h}x_0^{ - {\alpha _h}}{h_0}}}{{P_n + \sum\limits_{{\Phi _h}\backslash \left\{ {{o_h}} \right\}} {{P_h}x_i^{ - {\alpha _h}}{h_i}}  + \sum\limits_{{\Phi _g}} {{P_g}y_j^{ - {\alpha _g}}{g_j}} }} > \varepsilon } \right \}\notag\\
		&{\rm{ = }}{\rm{Pr}}\left \{ {{h_0} > \varepsilon x_0^{{\alpha _h}}( {\frac{P_n}{{{P_h}}} + \sum\limits_{{\Phi _h}\backslash \left\{ {{o_h}} \right\}} {x_i^{ - {\alpha _h}}{h_i}}  + \sum\limits_{{\Phi _g}} {\frac{{{P_g}}}{{{P_h}}}y_j^{ - {\alpha _g}}{g_j}} })} \right \}\vspace{1ex}\notag\\
		&\mathop {\rm{ = }}\limits^{(\text{a})} {\rm{E}}\left \{ {\frac{{\Gamma ( {m,m\varepsilon x_0^{{\alpha _h}}( {\frac{P_n}{{{P_h}}} + {I_{h1}} + {I_{g1}}} )})}}{{\Gamma \left( m \right)}}}\right \},
		\label{pr_h1}
	\end{align}
where (a) uses the CCDF in  \eqref{nakagami_CCDF} and ${\rm{E(*)}}$ means the expectation. When $m$ is an integer, substituting  \eqref{upper incomplete gamma} into \eqref{pr_h1}, \eqref{pr_h1} is further expanded to
	\begin{align}
			{{\Pr}_h}\left\{ {{\Upsilon _{h1}} > \varepsilon } \right\} = \exp \left( { - m\varepsilon x_0^{{\alpha _h}}\frac{{{P_n}}}{{{P_h}}}} \right)f\left( {m\varepsilon x_0^{{\alpha _h}}} \right),
		\label{Pr_h1_temp}
	\end{align}
where 
\begin{equation}
	f\left( s \right) = E\left\{ {\exp \left[ { - s({I_{h1}}+{I_{g1}})} \right]\sum\limits_{k = 0}^{m - 1} {\frac{{{s^k}}}{{k!}}{{( {\frac{{{P_n}}}{{{P_h}}} + {I_{h1}} + {I_{g1}}})^k}}} } \right\}.
	\label{fs}
\end{equation}
Using the binomial theorem ${(a + b)^m} = \sum\limits_{k = 0}^m {C_m^k} {a^k}{b^{m - k}}$, \eqref{fs} can be expanded to \eqref{fs_expand}.

The Laplace transform of ${I_{h1}}$  is 
\begin{align}
	{L_{{I_{h1}}}}\left( s \right)	&= {{\rm{E}}_{{\Phi _h},{h_i}}}\{ {\prod\limits_{{\Phi _h}\backslash \{ {{o_h}} \}} {\exp( { - sx_i^{ - {\alpha _h}}{h_i}} )} }  \} \notag\\
	&\mathop  = \limits^{(\text{a})} {{\rm{E}}_{{\Phi _h}}} \{ {\prod\limits_{{\Phi _h}\backslash \left\{ {{o_h}} \right\}} {{{\rm{E}}_{{h_i}}}\left[ {\exp \left( { - sx_i^{ - {\alpha _h}}{h_i}} \right)} \right]} } \},
	\label{Lh1_temp}	
\end{align}
 where (a) applies the expectation of $\exp( { - sx_i^{ - {\alpha _h}}{h_i}} )$ with respect to $h_i$ because of the independence of different air-to-ground channels. According to the PDF of air-to-ground channel power gain in \eqref{nakagami_PDF}, we have  \eqref{expection_hi}.
\begin{align}
		&{{\rm{E}}_{{h_i}}}\left\{ {\exp \left( { - sx_i^{ - {\alpha _h}}{h_i}} \right)} \right\}\notag \\
		&{\rm{ = }}\int_0^{ + \infty } {\exp \left( { - sx_i^{ - {\alpha _h}}{h_i}} \right)\frac{{{m^m}h_i^{m - 1}}}{{\Gamma \left( m \right)}}\exp \left( { - m{h_i}} \right)} d{h_i}\notag \\
		&= \frac{{{m^m}}}{{{{\left( {m + sx_i^{ - {\alpha _h}}} \right)}^m}}}.
		\label{expection_hi}
\end{align}
The PPP has the following probability generating functional \cite{haenggi2012stochastic}.
\begin{equation}
	{\rm{E}}( {\prod\limits_{x \in \Phi } {v\left( x \right)} }) = \exp \left ( { - \int_{{\mathbb{R}^d}} {\left[ {1 - v\left( x \right)} \right]\Lambda \left( {dx} \right)} } \right )
	\label{probability generating functional},
\end{equation}
where $\Phi$ is a PPP on Euclidean space $\mathbb{R}^d$  with intensity measure $\Lambda$  and $v$ is a measurable function on $\mathbb{R}^d$.
Then,  \eqref{Lh1_temp} is expressed as \eqref{Lh1} according to the equivalent density in \eqref{the equivalent density}.
\begin{align}
&{L_{{I_{h1}}}}\left( s \right) \notag\\
&= \exp \left( { - {\lambda _h}\int_0^{2\pi } {\int_0^t {[ {1 - \frac{{{m^m}}}{{{{\left( {m + sx_i^{ - {\alpha _h}}} \right)}^m}}}} ]{b^2}\sin \theta d\theta d\varphi } } } \right)
	\label{Lh1}
\end{align}
The $x_i$ in \eqref{Lh1} can be given by ${x_i} = \sqrt {{a^2} + {b^2} - 2ab\cos \theta }$. Similarly, the Laplace transform of $I_{g1}$ is
\begin{align}
		&{L_{{I_{g1}}}}\left( s \right) \notag\\
		&= {{\rm{E}}_{{\Phi _g}}}\left\{ {\prod\limits_{{\Phi _g}} {{{\rm{E}}_{{g_i}}}\left\{ {\exp \left( { - s\frac{{{P_g}}}{{{P_h}}}y_j^{ - {\alpha _g}}{g_i}} \right)} \right\}} } \right\}\notag \\
		& \mathop  = \limits^{{\rm{(a)}}} {{\rm{E}}_{{\Phi _h}}}\left\{ {\prod\limits_{{\Phi _g}} {\frac{1}{{1 + 2{\sigma ^2}s\frac{{{P_g}}}{{{P_h}}}y_j^{ - {\alpha _g}}}}} } \right\}\label{Lg1} \\ 
		& \mathop  = \limits^{{\rm{(b)}}} \exp \left( { - {\lambda _g}\int_0^{2\pi } {\int_0^{{r_b}} {\left[ {1 - \frac{1}{{1 + 2{\sigma ^2}s\frac{{{P_g}}}{{{P_h}}}y_{}^{ - {\alpha _g}}}}} \right]ydyd\varphi } } } \right),\notag
\end{align}
where (a) utilizes the PDF of ground-to-ground channel power gain in  \eqref{rayleigh_PDF} and (b) utilizes the probability generating functional in  \eqref{probability generating functional}.
Substituting $s=m \varepsilon x_0^{{\alpha _h}}$ into  \eqref{Lh1} and \eqref{Lg1}, we can get ${\rm{E}}[ {\exp ( { - m\varepsilon x_0^{{\alpha _h}}{I_{h1}}} )}]$ and ${\rm{E}}[ {\exp ( { - m\varepsilon x_0^{{\alpha _h}}{I_{g1}}} )}]$, respectively. According to the definition of $L_{I_{h1}}(s)$ and $L_{I_{g1}}(s)$, we have
\begin{equation}
	\frac{{{\partial ^z}}}{{\partial {s^z}}}{L_{{I_{h1}}}}\left( s \right){\rm{ = }}{{\rm{E}}_{{I_{h1}}}}\left\{ {{{\left( { - 1} \right)}^z}{{\left( {{I_{h1}}} \right)}^z}\exp \left( { - s{I_{h1}}} \right)} \right\},
	\label{D_Lh1}
\end{equation}
\begin{equation}
	\frac{{{\partial ^{k - l - z}}}}{{\partial {s^{k - l - z}}}}{L_{{I_{g1}}}}\left( s \right){\rm{ = }}{{\rm{E}}_{{I_{g1}}}}\left\{ {{{\left( { - 1} \right)}^{k - l - z}}{{\left( {{I_{g1}}} \right)}^{k - l - z}}\exp \left( { - s{I_{g1}}} \right)} \right\}
	\label{D_Lg1}.
\end{equation}
We can obtain the results of ${\rm{E}}[{\left( {{I_{h1}}} \right)^z}\exp ( - 2\varepsilon x_0^{{\alpha _h}}{I_{h1}})]$ and ${\rm{E}}[{\left( {{I_{g1}}} \right)^{k - l - z}}\exp ( - 2\varepsilon x_0^{{\alpha _h}}{I_{g1}})]$ by substituting $s=m \varepsilon x_0^{{\alpha _h}}$ into \eqref{D_Lh1} and \eqref{D_Lg1}, respectively.
Then, the coverage probability of HAP network can be derived as  \eqref{coverage_exact_h1}.
\end{proof}

In this paper, we mainly consider the case where $m$ is a positive integer, \cite{initial_Nakagami,nakagami_rician} mention that for non-integer cases where $m>1$, the Nakagami-m distribution with  shape parameter and scaling parameter of $(m,1/m)$ can be approximated using the Rician distribution with shape parameters of $(K,1)$, $K = m - 1 + \sqrt {m\left( {m - 1} \right)} $. According to \cite{k-u}, the PDF of power gain of Rician distribution can be approximated by the weighted sum of infinite Gamma distributions. In other words, the Nakagami distribution with $m$ taking a non-integer can be represented by the weighted sum of infinite Nakagami distributions with $m$ taking an integer. Therefore, it is fundamental to discuss the Nakagami distribution in which $m$ is an integer.

It is revealed from \eqref{coverage_exact_h1} that when $m>2$, it involves the higher-order derivative of the Laplace transform. The computational complexity is extremely high, which is difficult to be applied in the network optimization. To alleviate this problem, an inequality for the lower incomplete gamma function is given in \textbf{Lemma 1} as follows. It is worth noting that \cite{whynakagami} used this inequality for the first time to derive the coverage probability of vertical heterogeneous network. Then, we derive the approximate expression of \eqref{pr_h1} in \textbf{Theorem 2}.

\textbf{Lemma 1:} Denote  $\gamma \left( {m,mx} \right)  = \int_0^{mx} {{y^{m - 1}}{e^{ - y}}} dy$ as the lower incomplete gamma function. Then, the lower and upper bound of $\gamma \left( {m,mx} \right)$  are \cite{gamma1}
\begin{equation}
	{\left( {1 - {e^{ - {\eta _1}mx}}} \right)^m} \le \frac{{\gamma \left( {m,mx} \right)}}{{\Gamma \left( m \right)}} \le {\left( {1 - {e^{ - {\eta _2}mx}}} \right)^m}
	\label{The lower and upper bound of the lower incomplete Gamma function},
\end{equation}
where
\begin{equation}
	\centering
	{\eta _1}{\rm{ = }}\left\{ {\begin{array}{*{20}{c}}
			1&{m \ge 1}\\
			{{{\left( {m!} \right)}^{ - \frac{1}{m}}}}&{m \le 1}
	\end{array}} \right., 
	{\eta _2}{\rm{ = }}\left\{ {\begin{array}{*{20}{c}}
			{{{\left( {m!} \right)}^{ - \frac{1}{m}}}}&{m \ge 1}\\
			{\rm{1}}&{m \le 1}
	\end{array}} \right..
\end{equation}

\textbf{Theorem 2:}  The approximate expression of HAP network's coverage probability is \eqref{coverage_appro_h1}, 
\begin{figure*}[htb]
\begin{align}
	{{\rm{Pr}}_h}\left\{ {\varUpsilon_{h1} > \varepsilon } \right\}
	{\rm{ = }}\sum\limits_{k = 1}^m {C_m^k} {\left( { - 1} \right)^{k + 1}}\exp( { - k{\eta _2}m\varepsilon x_0^{{\alpha _h}}\frac{{{P_n}}}{{{P_h}}}}){\left. {{L_{{I_{h1}}}}\left( s \right){L_{{I_{g1}}}}\left( s \right)} \right|_{s = k{\eta _2}m\varepsilon x_0^{{\alpha _h}}}}.
	\label{coverage_appro_h1}
\end{align}	
\begin{equation}
	{L'_{{I_{h1}}}}\left( s \right) = \exp \left ( { - \lambda _h^*\int_0^{2\pi } {\int_0^t {\int_{b - \Delta h}^{b + \Delta h} {\left[ {{\rm{1 - }}\frac{m^m}{{{{\left( {m + s(\sqrt {{a^2} + {r^2} - 2ar\cos \theta })^{ - {\alpha _h}}} \right)}^m}}}} \right]{r^2}\sin \theta drd\theta d\varphi } } } } \right).
	\label{L_I_h1}
\end{equation}
{\noindent} \rule[-10pt]{18cm}{0.05em}
\end{figure*}
 where $C^k_m{\rm{ = }}\frac{{m!}}{{k!\left( {m - k} \right)!}}$, and $m!$, $k!$, $(m-k)!$ represent the factorial of $m$, $k$ and $m-k$, respectively.

\begin{proof}
	According to the definition of $\gamma \left( {m,mx} \right)$ in \textbf{Lemma 1}, the relation between $\Gamma \left( {m,mx} \right)$ and $\gamma \left( {m,mx} \right)$ is as follows
	\begin{equation}
		\frac{{\Gamma \left( {m,mx} \right)}}{{\Gamma \left( m \right)}}{\rm{ = }}1 - \frac{{\gamma \left( {m,mx} \right)}}{{\Gamma \left( m \right)}}.
	\end{equation}
	Then, the coverage probability of HAP network, i.e. \eqref{pr_h1} can be expressed as
	\begin{equation}
		{{\rm{Pr}}_h}\left\{ {\varUpsilon_{h1} > \varepsilon } \right\}
		{\rm{ = E}}\left \{ {1 - \frac{{\gamma \left ( {m,m\varepsilon x_0^{{\alpha _h}}( {\frac{P_n}{{{P_h}}} + {I_{h1}} + {I_{g1}}})} 
					\right )}}{{\Gamma \left( m \right)}}}\right \}.
		\label{appro_coverage_temp}
	\end{equation}

The upper bound in  \eqref{The lower and upper bound of the lower incomplete Gamma function}  is a good approximation of lower incomplete Gamma function, which has been confirmed in \cite{whynakagami} and \cite{gamma2}. Then, the approximation of  \eqref{appro_coverage_temp} is
\begin{align}
	&{{\rm{Pr}}_h}\left\{ {\varUpsilon_{h1} > \varepsilon } \right\} \notag \\
	&= 1 - {\rm{E}}\left\{ {{{\left[ {1 - \exp \left ( { - {\eta _2}m\varepsilon x_0^{{\alpha _h}}( {\frac{P_n}{{{P_h}}} + {I_{h1}} + {I_{g1}}} )} \right )} \right]^m}}}\label{approximation} \right\}.
\end{align}

According to the binomial theorem, the \eqref{approximation} can be further simplified as \eqref{coverage_appro_h1}, where $L_{I_{h1}}(s)$ and $L_{I_{g1}}(s)$ are the same as  \eqref{Lh1} and \eqref{Lg1}.
\end{proof}

\subsection{Fluctuation of HAP's height }
As HAPs are deployed in the stratosphere, they are susceptible to height fluctuation due to airflow. This section investigates the impact of height fluctuation on coverage probability of HAP network, assuming that HAPs fluctuate within a spherical shell around the deployment height. The coordinate of the HAP can be expressed in spherical coordinates as follows.
\begin{equation}
	\emph{\textbf{r}}\left( {r,\theta ,\varphi } \right){\rm{ = }}\left( {r\sin \theta \cos \varphi ,r\sin \theta \sin \varphi ,r\cos \theta } \right),
\end{equation}
where $b - \Delta h \le r \le b + \Delta h$, $0 \le \theta  \le t$, $0 \le \varphi  \le 2\pi $. $r$ is the distance between the HAP and the earth's center. $\Delta h$ is the height of HAP fluctuations. The distance between the typical HAP user and associated HAP is
\begin{equation}
	{x_i} = \sqrt {{a^2} + {r^2} - 2ar\cos \theta }.
\end{equation}

The equivalent distribution density is
\begin{equation}
	\lambda _h^* = \frac{E(N)}{{2\Delta hdS}} = \frac{{{\lambda _h}}}{{2\Delta h}},
	\label{equal_lambda_1}
\end{equation}
 where $dS$ is the area of surface, and $E(N)$ is the average number of points on this surface. Then, the coverage probability of HAP network when considering fluctuation of HAP's height is provided in \textbf{Theorem 3}.
 
\textbf{Theorem 3:}  The exact and approximate expressions of the coverage probability of HAP network when considering fluctuation of HAP's height are the expectation of \eqref{coverage_exact_h1} and \eqref{coverage_appro_h1} with respect to $x_0$, where  the expression of $L_{I_{h1}}(s)$ is replaced by \eqref{L_I_h1}.

\begin{proof}
	The impact of HAP fluctuation on typical HAP user is mainly determined in the aggregated interference from other HAPs. Therefore, considering the fluctuation of HAP's height, the expressions of the coverage probability are only different in $L_{I_{h1}}(s)$. Through the probability generating functional in  \eqref{probability generating functional} and equivalent distribution density in  \eqref{equal_lambda_1}, the $L_{I_{h1}}(s)$ in  \eqref{Lh1_temp} can be derived as  \eqref{L_I_h1}.
\end{proof}

Generally speaking, the fluctuation range of HAP is much smaller than the radius of the spherical surface where it is located, i.e. $\Delta h \ll b$. Then, we have	
\begin{equation}
	\begin{array}{l}
		\int_{b - \Delta h}^{b + \Delta h} {[1 - \frac{{{m^m}}}{{{{(m + s{{(\sqrt {{a^2} + {r^2} - 2ar\cos \theta } )}^{ - {\alpha _h}}})}^m}}}]{r^2}\sin \theta dr} \\
		\approx 2\Delta h[1 - \frac{{{m^m}}}{{{{(m + s{{(\sqrt {{a^2} + {b^2} - 2ar\cos \theta } )}^{ - {\alpha _h}}})}^m}}}]{b^2}\sin \theta .
	\end{array}
\end{equation}
Comparing \eqref{Lh} and \eqref{L_I_h1}, we have  ${L'_{{I_{h1}}}}\left( s \right) \approx {L_{{I_{h1}}}}\left( s \right)$. Therefore, the fluctuation of HAP's height has a little impact on coverage probability of HAP network.	

\subsection{Coverage probability of terrestrial network}
Since  HAP network shares spectrum with terrestrial network, the HAP downlink will cause interference to the downlink of the terrestrial network. In order to study the impact of the interference from HAP network on the performance of terrestrial network, it is necessary to derive the coverage probability of terrestrial network.

For the terrestrial network, the signal received by a typical BS user is interfered by the HAPs and neighboring BSs, with SINR as follows.
\begin{equation}
	\varUpsilon_{g1} = \frac{{{P_g}y_0^{ - {\alpha _g}}{g_0}}}{{P_n + \sum\limits_{{x_i} \in {\Phi _h}} {{P_h}x_i^{ - {\alpha _h}}{h_i} + \sum\limits_{{y_j} \in {\Phi _g}\setminus\left\{ {{o_b}} \right\}} {{P_g}y_j^{ - {\alpha _g}}{g_j}} } }},
	\label{sinr_bs}
\end{equation}
where $y_0$ is the distance between the typical BS user and associated BS. Then, the coverage probability of terrestrial network is obtained in \textbf{Theorem 4}.

\textbf{Theorem 4:} The coverage probability of terrestrial network is
\begin{equation}
	{{\rm{Pr}}_g}\{ {\varUpsilon_{g1} > \varepsilon }\} = \exp ( { - \frac{\varepsilon y_0^{{\alpha _g}}}{2\sigma^2}\frac{P_n}{{{P_g}}}} ){L_{{I_{{h_2}}}}}( \frac{\varepsilon y_0^{{\alpha _g}}}{2\sigma^2} ){L_{{I_{{g_2}}}}}( \frac{\varepsilon y_0^{{\alpha _g}}}{2\sigma^2}),
	\label{coverage_BS}
\end{equation}
where ${I_{h2}} = \sum\nolimits_{{x_i} \in {\Phi _h}} {x_i^{ - {\alpha _h}}{h_i}{P_h}/{P_g}}$ and  ${I_{g2}} = \sum\nolimits_{{y_j} \in {\Phi _g}\setminus\left\{ {{o_b}} \right\}} {y_j^{ - {\alpha _g}}{g_j}}$ represent the aggregated interference from HAPs and other BSs, respectively. The expressions of ${L_{{I_{h2}}}}( \frac{\varepsilon y_0^{{\alpha _g}}}{2\sigma^2} )$  and  ${L_{{I_{g2}}}}( \frac{\varepsilon y_0^{{\alpha _g}}}{2\sigma^2} )$ can be obtained by substituting $s= \frac{\varepsilon y_0^{{\alpha _g}}}{2\sigma^2}$ into  \eqref{L_h2} and \eqref{L_g2}.
\begin{figure*}
	\begin{equation}
		{L_{{I_{h2}}}}(s) = \exp \left ( { - \int_0^{2\pi } {\int_0^t {\left[ {1 - \frac{m^m}{{{{\left( {m + \frac{{{P_h}}}{{{P_g}}}s(\sqrt {{a^2} + {b^2} - 2ab\cos \theta })^{ - {\alpha _h}}} \right)}^m}}}} \right]{\lambda _h}{b^2}\sin \theta d\theta d\varphi } } } \right ).
		\label{L_h2}	
	\end{equation}
\end{figure*}
\begin{equation}
	{L_{{I_{g2}}}}(s) = \exp \left ( { - \int_0^{2\pi } {\int_0^{r_b} {\left[ {1 - \frac{1}{{1 + sy_{}^{ - {\alpha _g}}}}} \right]} } {\lambda _g}ydyd\varphi } \right ).
	\label{L_g2}
\end{equation}

\begin{proof}
	According to \eqref{sinr_bs}, the coverage probability of terrestrial network can be expressed as
	\begin{align}
		&{\text{Pr}_g}\left\{{\varUpsilon_{g1} > \varepsilon } \right) \notag \\
		&= {\text{Pr} _g}\left\{ {\frac{{{P_g}y_0^{ - {\alpha _g}}{g_0}}}{{P_n + \sum\limits_{{x_i} \in {\Phi _h}} {{P_h}x_i^{ - {\alpha _h}}{h_i} + \sum\limits_{{y_j} \in {\Phi _g}\setminus\left\{ {{o_b}} \right\}} {{P_g}y_j^{ - {\alpha _g}}{g_j}} } }} > \varepsilon } \right\}\notag\\
		&= {\text{Pr}_g}\left\{ {{g_0} > \varepsilon y_0^{{\alpha _g}} ( {\frac{{{P_n}}}{{{P_g}}} + I_{h2}+ I_{g2} } )} \right\}	\label{Pr_g2}\\
		&\mathop {\rm{ = }}\limits^{\left( {\rm{a}} \right)} \exp ( { - \frac{\varepsilon y_0^{{\alpha _g}}}{2\sigma^2}\frac{P_n}{{{P_g}}}} ){L_{{I_{{h_2}}}}}( \frac{\varepsilon y_0^{{\alpha _g}}}{2\sigma^2} ){L_{{I_{{g_2}}}}}( \frac{\varepsilon y_0^{{\alpha _g}}}{2\sigma^2}),\notag
	\end{align}
 where (a) utilizes the CCDF of ground-to-ground channel power gain in  \eqref{rayleigh_CCDF}. Similar to the proof  for \eqref{Lh1} and \eqref{Lg1}, we can derive the expressions of ${L_{{I_{h2}}}}( s )$  and  ${L_{{I_{g2}}}}( s )$ as  \eqref{L_h2} and \eqref{L_g2}.
\end{proof}

It is revealed from \eqref{coverage_BS} and \eqref{L_h2} that as the HAP density increases, the coverage probability of the terrestrial network decreases. The distance from BS to the typical user is much smaller than the distance from HAP to the typical user, so that  $y_0^{{\alpha _g}}{(\sqrt {{a^2} + {b^2} - 2ab\cos \theta } )^{ - {\alpha _h}}} \approx 0$. Then we can get ${L_{{I_{h2}}}}(\frac{{\varepsilon y_0^{{\alpha _g}}}}{{2{\sigma ^2}}}) \approx 1$ from \eqref{L_h2} and ${\rm{P}}{{\rm{r}}_g}\{ {\Upsilon _{g1}} > \varepsilon \}  \approx \exp ( - \frac{{\varepsilon y_0^{{\alpha _g}}}}{{2{\sigma ^2}}}\frac{{{P_n}}}{{{P_g}}}){L_{{I_{{g_2}}}}}(\frac{{\varepsilon y_0^{{\alpha _g}}}}{{2{\sigma ^2}}})$ from \eqref{coverage_BS}. Thus, the HAP network has limited interference to the terrestrial network.

\subsection{Transmission capacity of HAP network}
The transmission capacity of wireless network is defined as the number of successful transmissions per unit area, constrained by the outage probability \cite{TC1}. According to \cite{PPP3}, the transmission capacity of HAP network is
\begin{equation}
	T_C = {\lambda _h}{{\Pr }_h}\left\{ {\varUpsilon_h > \varepsilon } \right\}\log (1 + \varepsilon ).
	\label{TC}
\end{equation}
There are no HAPs when ${\lambda _h}=0$, so that the transmission capacity is 0. It can be revealed from \eqref{TC} that the transmission capacity is greater than 0 when ${\lambda _h}>0$. The increase of HAP density will increase the interference to air-to-ground communication, deteriorating the coverage probability. The transmission capacity tends to 0 when ${\lambda _h} \to \infty $. Because the coverage probability of HAP network is the high-order infinitesimal of ${\lambda _h}$, which can be derived  from \eqref{coverage_appro_h1}. Therefore, there is an optimal density to maximize the transmission capacity.

Taking the derivative of $T_C$ with respect to $\lambda_h$ yields \eqref{derive_Tc}, where ${\Pr _h}\left( k \right)$ is given in \eqref{Pr_k}. By making \eqref{derive_Tc} equal to 0, the optimal density can be obtained as shown in \eqref{opt_density}. 
\begin{figure*}[htb]
	\vspace{-0.5cm}
	\begin{equation}
		\frac{\partial }{{\partial {\lambda _h}}}{T_C} = \log \left( {1 + \varepsilon } \right)\sum\limits_{k = 1}^m {\left[ {1 - 2\pi {\lambda _h}\int_0^t {[1 - \frac{1}{{{{\left( {1 + k{\eta _2}\varepsilon x_0^{{\alpha _h}}{{(\sqrt {{a^2} + {b^2} - 2ab\cos \theta } )}^{ - {\alpha _h}}}} \right)}^m}}}]{b^2}\sin \theta d\theta } } \right]{{\Pr }_h}\left( k \right)}.
		\label{derive_Tc}
	\end{equation}
	\begin{equation}
		{\text{Pr}_h}\left( k \right) = C_m^k{\left( { - 1} \right)^{k + 1}}\exp \left( { - k{\eta _2}m\varepsilon x_0^{{\alpha _h}}\frac{{{P_n}}}{{{P_h}}}} \right){L_{{I_{g1}}}}\left( {k{\eta _2}m\varepsilon x_0^{{\alpha _h}}} \right){L_{{I_{h1}}}}\left( {k{\eta _2}m\varepsilon x_0^{{\alpha _h}}} \right).
		\label{Pr_k}
	\end{equation}
	\begin{equation}
		\lambda _h^{opt}{\rm{ = }}\frac{{\sum\limits_{k = 1}^m {{{\Pr }_h}\left( k \right)} }}{{2\pi \sum\limits_{k = 1}^m {\int_0^t {\left[ {1 - \frac{1}{{{{\left( {1 + k{\eta _2}\varepsilon x_0^{{\alpha _h}}{{(\sqrt {{a^2} + {b^2} - 2ab\cos \theta } )}^{ - {\alpha _h}}}} \right)}^m}}}} \right]{b^2}\sin \theta d\theta } {{\Pr }_h}\left( k \right)} }}.
		\label{opt_density}
	\end{equation}
	\begin{equation}
		\frac{1}{{2\pi \left[ {1 - \frac{1}{{{{\left( {1 + m{\eta _2}\varepsilon } \right)}^m}}}} \right]h\left( {h + a} \right)}} < \lambda _h^{opt}{\rm{ < }}\frac{1}{{2\pi \left[ {1 - \frac{{{{\left( {\sqrt {b + a} } \right)}^{m{\alpha _h}}}}}{{{{\left( {{{\left( {\sqrt {b + a} } \right)}^{{\alpha _h}}} + {\eta _2}\varepsilon {{(\sqrt {b - a} )}^{{\alpha _h}}}} \right)}^m}}}} \right]h\left( {h + a} \right)}}.
		\label{upper_lower_density}
	\end{equation}
{\noindent} \rule[-10pt]{18cm}{0.05em}
\end{figure*}

However, it is difficult to obtain the closed-form solution of the optimal density, so that we derived its upper and lower bounds. Because $\cos t \le \cos \theta  \le 1$, combining \eqref{opt_density} and \eqref{expression_t}, the upper and lower bounds of the optimal density can be obtained as \eqref{upper_lower_density}. From \eqref{upper_lower_density}, it is revealed that as the HAP height increases, the optimal density decreases.

The following optimization problem can be solved to find the deployment density that maximizes the transmission capacity of  HAP network while satisfying the constraint $\kappa$ of the coverage probability of terrestrial network.
\begin{equation}
	\begin{array}{*{20}{c}}
		{\mathop {\max }\limits_{{\lambda _h}} }&{T_C}\\
		{s.t.}&{{{\Pr }_g}\left\{ {\varUpsilon_g > \varepsilon } \right\} > \kappa },
	\end{array}
	\label{optimal}
\end{equation}
However, it's difficult to obtain the closed-form solution of this optimization problem. Therefore, we solve it numerically in Section VI.

\section{HAP Network with Directional Antenna}
As HAP cannot use fully dynamic beamforming technology because of limited payload 
\cite{direc_antenna}, we explore the application of directional antenna with fixed antenna pattern in HAP network.  Aiming to verify the performance improvement of directional antenna for spectrum sharing between HAP network and terrestrial network, the coverage probabilities of HAP network and terrestrial network are derived. 

\subsection{Coverage probability of HAP network}
\begin{figure}[htb]
	\centering
	\includegraphics[scale=0.55]{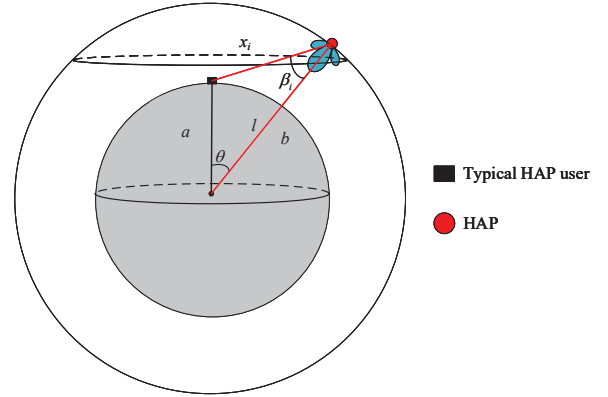}
	\caption{\centering { Distribution model of BSs and HAPs with directional antenna.}}
	\label{Figure_3}
\end{figure}

To improve the performance of HAP network, the directional antenna can be mounted on HAP. Without considering the inclination of HAP, the main beam is perpendicular to the ground and the center line $l$ of main beam is over the earth's center. It is revealed in Fig. \ref{Figure_3} that the angle ${\beta_i}$  between the downlink of HAP and the line $l$ is obtained according to the geometric relation, as shown in  \eqref{beta_i}. 
\begin{align}
	{\beta _i} = {\cos ^{ - 1}}( {\frac{{x_i^2 + {b^2} - {a^2}}}{{2b{x_i}}}} ){\rm{ = }}{\cos ^{ - 1}}( {\frac{{b - a\cos \theta }}{{\sqrt {{a^2} + {b^2} - 2ab\cos \theta } }}}).
	\label{beta_i}
\end{align}

Then, the SINR at typical HAP user is
\begin{equation}
	\varUpsilon_{h2} = \frac{{{P_h}G\left( {{\beta _0}} \right)x_0^{ - {\alpha _h}}{h_0}}}{{P_n + \sum\limits_{{x_i} \in \Phi_h \setminus\left\{ {{o_h}} \right\}} {{P_h}G\left( {{\beta _i}} \right)x_i^{ - {\alpha _h}}{h_i} + \sum\limits_{{y_j} \in {\Phi _g}} {{P_g}y_j^{ - {\alpha _g}}{g_j}} } }},
	\label{sinr}
\end{equation}
where ${\beta _0}$  is the angle between the downlink of HAP associated with the typical HAP user and the center line $l$.

Similarly, the coverage probability of HAP network mounting directional antenna can be derived as \textbf{Theorem 5}.

\begin{figure*}
	\vspace{1cm}
	\begin{align}
		{{\Pr }_h}\left\{ {{\varUpsilon _{h2}} > \varepsilon } \right\} = \exp ( - m\varepsilon x_0^{{\alpha _h}}\frac{{{P_n}}}{{{P_h}G\left( {{\beta _0}} \right)}})
		{\left. {\left\{ {\sum\limits_{k = 0}^{m - 1} {\frac{{{s^k}}}{{k!}}\sum\limits_{l = 0}^k {C_k^l{{\left( {\frac{{{P_n}}}{{{P_h}G\left( {{\beta _0}} \right)}}} \right)}^l}\sum\limits_{z = 0}^{k - l} {{{( - 1)}^{k - l}}C_{k - l}^zL_{{I_{h3}}}^{(z)}(s)L_{{I_{g3}}}^{k - l - z}(s)} } } } \right\}} \right|_{s = m\varepsilon x_0^{{\alpha _h}}}}
		\label{pr_exact_h3}.
	\end{align}
	\begin{equation}
		{{\rm{Pr}}_h}\left\{ {\varUpsilon_{h2} > \varepsilon } \right\} \approx \sum\limits_{k = 1}^m C^k_m {( { - 1} )^{k + 1}}\exp ( { - k\eta_2 m\varepsilon x_0^{{\alpha _h}}\frac{P_n}{{{P_h}G\left( {{\beta _0}} \right)}}} ){L_{{I_{h3}}}}\left( {k\eta_2 m\varepsilon x_0^{{\alpha _h}}} \right){L_{{I_{g3}}}}\left( {k\eta_2 m\varepsilon x_0^{{\alpha _h}}} \right)
		\label{pr_appro_h3}.
	\end{equation}
	\begin{equation}
		{L_{{I_{h3}}}}\left( s \right) = \exp \left( { - {\lambda _h}\int_0^{2\pi } {\int_0^t {[ {1 - \frac{m^m}{{{{( {m + s\frac{{G\left( \beta  \right)}}{{G\left( {{\beta _0}} \right)}}{{(\sqrt {{a^2} + {b^2} - 2ab\cos \theta } )}^{ - {\alpha _h}}}} )}^m}}}} ]} } {b^2}\sin \theta d\theta d\varphi } \right).
		\label{L_h3}
	\end{equation}
	\begin{equation}
		{L_{{I_{g3}}}}\left( s \right) = \exp \left ( { - {\lambda _g}\int_0^{2\pi } {\int_0^{r_b} {\left[ {1 - \frac{1}{{1 + {2{\sigma ^2}}s\frac{{{P_{\rm{g}}}}}{{{P_h}G\left( {{\beta _0}} \right)}}y^{ - {\alpha _g}}}}} \right]} } ydyd\varphi } \right).
		\label{L_g3}
	\end{equation}
	\begin{equation}
		{L'_{{I_{h3}}}}\left( s \right) = \exp  \left ( { - \lambda _h^*\int_0^{2\pi } {\int_0^t {\int_{b - \Delta h}^{b + \Delta h} {{[ {1 - \frac{m^m}{{{{( {m + s\frac{{G\left( \beta  \right)}}{{G\left( {{\beta _0}} \right)}}{{(\sqrt {{a^2} + {r^2} - 2ar\cos \theta } )}^{ - {\alpha _h}}}} )}^m}}}} ]}} } {r^2}\sin \theta dr d\theta d\varphi } } \right ).
		\label{L_h3_new}
	\end{equation}
{\noindent} \rule[-10pt]{18cm}{0.05em}
\end{figure*} 
\textbf{Theorem 5:} The exact and approximate expressions of the coverage probability of HAP network with directional antenna are derived as \eqref{pr_exact_h3} and \eqref{pr_appro_h3}, where  ${I_{h3}} = \sum\nolimits_{{\Phi _h}\backslash \left\{ {{o_h}} \right\}} {G\left( {{\beta _i}} \right)x_i^{{\alpha _h}}{h_i}} /G\left( {{\beta _0}} \right)$ and ${I_{g3}} = \sum\nolimits_{{\Phi _g}} {{P_g}y_j^{ - {\alpha _g}}{g_j}} /\left( {{P_h}G\left( {{\beta _0}} \right)} \right)$ represent the aggregated interference from other HAPs and terrestrial BSs, respectively. And the  Laplace transform of ${I_{h3}}$ and ${I_{g3}}$  can be found in \eqref{L_h3} and \eqref{L_g3}. 

\begin{proof}
	Comparing  \eqref{sinr_h1} and \eqref{sinr}, the main difference in SINR is the directional antenna gain. 
	Similar to the derivation of  \eqref{coverage_exact_h1} and  \eqref{coverage_appro_h1}, the exact and approximate expressions of coverage probability of HAP network are derived as \eqref{pr_exact_h3} and \eqref{pr_appro_h3}. And the Laplace transformations of $I_{h3}$ and $I_{g3}$ are derived in \eqref{L_h3} and \eqref{L_g3} using the probability generating functional of PPP in \eqref{probability generating functional}. 
	For the antenna gain $G(\beta_i)$, the $\beta_i$ is derived  according to the geometric relation, i.e.  \eqref{beta_i}. Then, $\beta_i$ is substituted into the antenna pattern to get $G(\beta_i)$.
\end{proof}	
	
\subsection{Fluctuation of HAP's height}
 The coverage probability of HAP network with directional antenna when considering the fluctuation of HAPs are provided in \textbf{Theorem 6}.
 
\textbf{Theorem 6:}  The exact and approximate expressions of the coverage probability of HAP network when considering fluctuation of HAPs are the expectations of \eqref{pr_exact_h3} and \eqref{pr_appro_h3}, with the expression of $L_{I_{h3}}$ replaced by \eqref{L_h3_new}.

\begin{proof}
	The proof of the coverage probability of HAP network is similar to the derivation of \eqref{pr_exact_h3} and \eqref{pr_appro_h3}, the derivation details of which are ignored.
	Considering that HAPs fluctuate within a spherical shell around the deployment height, the equivalent HAP distribution density is
	\begin{equation}
		\lambda _h^* = \frac{E(N)}{{2\Delta hdS }} = \frac{{{\lambda _h}}}{{2\Delta h}}.
		\label{equal_lambda}
	\end{equation}
	Then we can derive the expression of $L_{I_{h3}}$ as  \eqref{L_h3_new} using probability generating functional of PPP in \eqref{probability generating functional}.
\end{proof}

\subsection{Coverage probability of terrestrial network}
When mounting the directional antenna on HAPs, the communication between typical BS user and its associated BS suffers from the interference from other BSs and the side beams of HAPs. The SINR at typical BS user is
\begin{equation}
	\varUpsilon_{g2} = \frac{{{P_g}y_0^{ - {\alpha _g}}{g_0}}}{{P_n + \sum\limits_{{x_i} \in {\Phi _h}} {{P_h}G\left( {{\beta _i}} \right)x_i^{ - {\alpha _h}}{h_i} + \sum\limits_{{y_j} \in {\Phi _g}\setminus\left\{ {{o_b}} \right\}} {{P_g}y_j^{ - {\alpha _g}}{g_j}} } }}.
\end{equation}

Then, the coverage probability of terrestrial network can be derived as \textbf{Theorem 7}.

\textbf{Theorem 7:}  The coverage probability of terrestrial network is \eqref{pr_g4} when  mounting the directional antenna on HAPs.
\begin{equation}
	{{\rm{Pr}}_g}\left\{ {\varUpsilon_{g2} > \varepsilon } \right\} = \exp ( { - \frac{\varepsilon y_0^{{\alpha _g}}}{2\sigma^2}\frac{P_n}{{{P_g}}}} ){L_{{I_{h4}}}}( { - \frac{\varepsilon y_0^{{\alpha _g}}}{2\sigma^2}} ){L_{{I_{g4}}}}( { - \frac{\varepsilon y_0^{{\alpha _g}}}{2\sigma^2}} ),
	\label{pr_g4}
\end{equation}
where ${I_{h4}} = \sum\nolimits_{{x_i} \in {\Phi _h}} {G\left( {{\beta _i}} \right)x_i^{ - {\alpha _h}}{h_i}{P_h}/{P_g}}$  denotes the aggregated interference from the side beams of HAPs and ${I_{g4}} = \sum\nolimits_{{y_j} \in {\Phi _g}\setminus\left\{ {{o_b}} \right\}} {y_j^{ - {\alpha _g}}{g_j}}$  denotes the aggregated interference from other BSs. The Laplace transforms of $I_{h4}$ and $I_{g4}$ are \eqref{L_h4} and \eqref{L_g4}, respectively.

\begin{proof}
	The proof is similar to that of \textbf{Theorem 4}. With the antenna gain added in \eqref{L_h4}, \eqref{pr_g4} can be derived. 
\end{proof}

Since the HAPs are implemented  with directional antennas, most of the interference of HAPs to the typical user of terrestrial network comes from the side lobes of HAPs with small gain, namely $G\left( \beta  \right) < 1$. Comparing \eqref{L_h2} and \eqref{L_h4}, we can get  ${L_{{I_{h4}}}}\left( {\frac{{\varepsilon y_0^{{\alpha _g}}}}{{2{\sigma ^2}}}} \right) > {L_{{I_{h2}}}}\left( {\frac{{\varepsilon y_0^{{\alpha _g}}}}{{2{\sigma ^2}}}} \right)$. Therefore, the coverage probability of terrestrial network is greater when the HAPs are equipped with directional antennas  compared with the case that the HAPs are equipped with omnidirectional antennas.	

\subsection{Transmission capacity of HAP network}
Similar to Section IV-D, we can get the upper and lower bounds of the optimal density for transmission capacity of HAP network with directional antenna as shown in \eqref{upper_lower_density_direc}.

\section{Simulation Results and Analysis}
The simulation parameters are listed in TABLE \ref{table_2}.  The generation of homogeneous PPP on the spherical surface is listed in TABLE \ref{table_3}. The antenna gain with directional antenna is as follows \cite{antenna2}, which has a main lobe of Gaussian form and constant level of side lobes.
\begin{equation}
	\begin{footnotesize}
	\begin{array}{l}
		G\left( {\beta ,{\theta _{ - 3dB}}} \right)[dB]{\rm{ = }}\left\{ {\begin{array}{*{20}{c}}
				{{G_0} - 3.01\left( {\frac{{2\beta }}{{{\theta _{ - 3dB}}}}} \right)}&{0 \le \beta  \le {\theta _{ml}}/2}\\
				{{G_{sl}}}&{{\theta _{ml}}/2 \le \beta  \le {{180}^o}}
		\end{array}} \right.{\rm{      }}\\
		{G_0} = 10\log \left( {\frac{{1.6162}}{{\sin \left( {{\theta _{ - 3dB}}/2} \right)}}} \right)\\
		{\theta _{ml}} = 2.6{\theta _{ - 3dB}}\\
		{G_{sl}} =  - 0.4111 \cdot \ln \left( {{\theta _{ - 3dB}}} \right) - 10.597
	\end{array}
	\label{antenna_model}
	\end{footnotesize}
\end{equation}
where $\beta$ is the angle between the communication link and the centerline of the main beam. ${\theta _{ - 3dB}}$ and ${\theta _{ml}}$ represent the half-power bandwidth and main lobe bandwidth of the antenna, respectively. $G_{sl}$ represents the antenna gain of side lobe.
\begin{figure*}
	\begin{equation}
		{L_{{I_{h4}}}}\left( s \right) = \exp \left( { - \int_0^{2\pi } {\int_0^t {\left[ {1 - \frac{m^m}{{{{\left( {m + \frac{{{P_h}}}{{{P_g}}}G\left( {{\beta}} \right)s(\sqrt {{a^2} + {b^2} - 2ab\cos \theta })^{ - {\alpha _h}}} \right)}^m}}}} \right]{\lambda _h}{b^2}\sin \theta d\theta d\varphi } } } \right).
		\label{L_h4}
	\end{equation}
	\begin{equation}
		{L_{{I_{g4}}}}\left( s \right) = \exp \left( { - \int_0^{2\pi } {\int_0^{r_b} {\left[ {1 - \frac{1}{{1 + sy_{}^{ - {\alpha _g}}}}} \right]} } {\lambda _g}ydyd\varphi } \right).
		\label{L_g4}
	\end{equation}
	\begin{equation}
		\frac{1}{{2\pi \left[ {1 - \frac{1}{{{{\left( {1 + m{\eta _2}\varepsilon \frac{{G\left( \beta  \right)}}{{G\left( {{\beta _0}} \right)}}} \right)}^m}}}} \right]h\left( {h + a} \right)}} < \lambda _h^{opt} < \frac{1}{{2\pi \left[ {1 - \frac{{{{\left( {\sqrt {b + a} } \right)}^{m{\alpha _h}}}}}{{{{\left( {{{\left( {\sqrt {b + a} } \right)}^{{\alpha _h}}} + {\eta _2}\varepsilon \frac{{G\left( \beta  \right)}}{{G\left( {{\beta _0}} \right)}}{{(\sqrt {b - a} )}^{{\alpha _h}}}} \right)}^m}}}} \right]h\left( {h + a} \right)}}.
		\label{upper_lower_density_direc}
	\end{equation}
	{\noindent} \rule[-10pt]{18cm}{0.05em}
\end{figure*}

\begin{table}[h]
	\caption{\label{sys_para}Simulation parameters}
	\begin{center}
		\begin{tabular}{c l c l}
			\hline
			\hline
			
			{Parameter} & {Value} &
			{Parameter} & {Value}\\
			
			\hline
			
			${P_h}$ & $50$ W&
			${P_g}$ & $20$ W\\
			${P_n}$ & $10^{-9}$  W&
			${\alpha_h}$ & $2$\\
			${\alpha_g}$ & $4$&
			${\lambda_h}$ & $1 \times 10^{-10}$ $\mbox{pieces/m}^2$ ($\mbox{pcs/m}^2$) \\
			${\lambda_g}$ & $3 \times 10^{-5}$ $\mbox{pcs/m}^2$ &
			$r_b$ & $10$ km\\
			$\epsilon$ & $0.1$ &
			$a$ & $6371$ km\\
			$h$ & $20-50$ km&
			$y_0$ & $50$ m\\
			$m$ & $2$&
			$\sigma^2$ & $0.5$ \\
			$\theta_{-3dB}$ & $40^{\circ}$\\
			\hline
			\hline
		\end{tabular}
	\end{center}
	\label{table_2}
\end{table}

\begin{table}[h]
	\caption{\label{sys_para}Generation of homogeneous PPP on spherical surface}
	\begin{center}
		\begin{tabular}{l p{8cm}}
			\hline
			\hline
			\multicolumn{2}{l}{\emph{Input}:  ${\lambda _h}$, $b$, ${\theta _{\min }}$, ${\theta _{\max }}$, ${\varphi _{\min }}$, ${\varphi _{\max }}$, number of intervals $N_i$.} \\
			\multicolumn{2}{l}{\emph{Output}: Homogeneous PPP ${\Phi _h}$  on spherical surface.} \\
			1.& Divide intervals $\left[ {{\theta _{\min }},{\theta _{\max }}} \right]$  and  $\left[ {{\varphi _{\min }},{\varphi _{\max }}} \right]$ evenly into $N_i$  subintervals to form $N_i^2$  disjoint sub-regions.\\
			2.& According to \eqref{the equivalent density}, the equivalent density $\lambda _h^*\left( \theta  \right)$  of the PPP on the $\theta  - \varphi $ plane is obtained.\\
			3. & Generate homogeneous PPP according to the maximum density ${\lambda _{\max }}$  in the sub-region and get the coordinates $\left( {\theta ,\varphi } \right)$ of each point.\\
			4. & Delete each point generated on the sub-region according to the probability $1 - \lambda _h^*\left( \theta  \right)/{\lambda _{\max }}$.\\
			5. & Repeat 3 and 4 until all sub-regions are scattered.\\
			6. & Map the points generated on the plane to the points on the spherical surface by \eqref{map} to get  ${\Phi _h}$.\\	
			\hline
			\hline
		\end{tabular}
	\end{center}
	\label{table_3}
\end{table}

\begin{figure}
	\centering
	\includegraphics[scale=0.52]{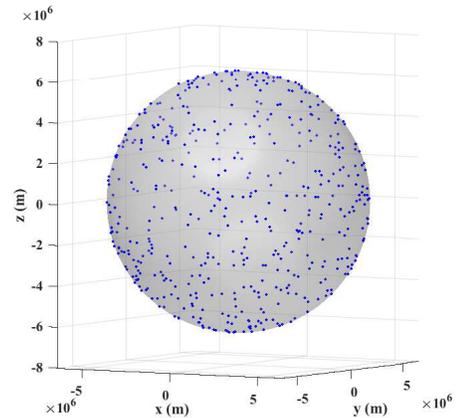}
	\caption{\centering {Generation of homogeneous PPP on a sphere.}}
	\label{figure_4}
\end{figure}

Fig. \ref{figure_4} shows a schematic diagram of homogeneous PPP on a sphere, generated by the method proposed in TABLE \ref{table_3}. In Monte Carlo simulation, the homogeneous PPP on sphere is converted to a non-homogeneous PPP on plane. Then the generated points are mapped onto the sphere.

\begin{figure}
	\centering
	\includegraphics[scale=0.52]{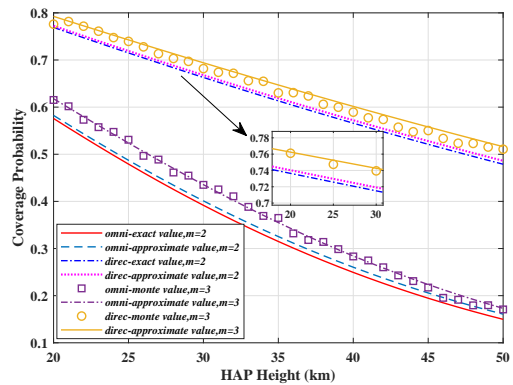}
	\caption{\centering {Comparison of exact and approximate values of the coverage probability of HAP.}}
	\label{figure_5}
\end{figure}

Fig. \ref{figure_5} shows the exact and approximate coverage probability of HAP when $m=2$, which reveals a small gap between the exact and approximate results. However, since it is difficult to derive the expression of exact value when $m>2$, the approximate coverage probability and the Monte Carlo value are shown when $m=3$. There is only a small gap between the exact and approximate results. Since the approximate value has a more concise expression and lower computational complexity, it is used as the theoretical values in the subsequent simulation.


\begin{figure}[htbp]
	\centering
	\hspace{-1cm}
	\includegraphics[scale=0.52]{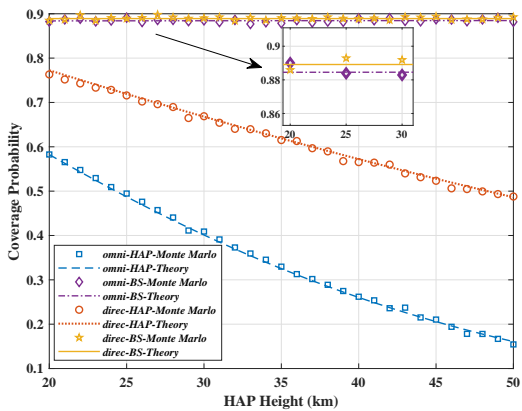}
	\caption{\centering {Coverage probability of HAP network and terrestrial network vs. HAP height.}}
	\label{figure_6}
\end{figure}

Fig. \ref{figure_6} shows the theoretical values of the coverage probability of HAP network and terrestrial network, while 10,000 Monte Carlo simulations are performed. It is revealed from  Fig. \ref{figure_6} that the theoretical derived values fit well with the Monte Carlo simulation values with different antenna conditions, thus verifying the accuracy of the theoretical derivation. The coverage probability of HAP decreases with the increase of the deployment height, which is due to the decrease of desired signal power received by typical HAP user with the increasing height. The application of directional antenna significantly improves the coverage probability of HAP network. Meanwhile, the coverage probability of terrestrial network has been maintained at a high level, which indicates that HAP network has less influence on terrestrial network with  spectrum sharing. On the other hand, when the HAPs are implemented with directional antenna, there is a small improvement in the coverage probability of terrestrial network, which is due to the fact that the interference at this situation comes from the side beams of HAPs and is smaller than the interference with omnidirectional antenna.

\begin{figure}[htbp]
	\centering
	\includegraphics[scale=0.52]{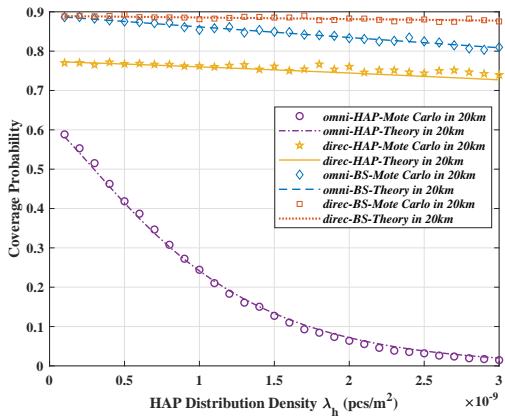}
	\caption{\centering {Coverage probability of HAP network and terrestrial network vs. HAP  density.}}
	\label{figure_7}
\end{figure}

Fig. \ref{figure_7} shows the coverage probability of HAP network and terrestrial network versus the deployment density with the deployment height at 20 km. It is revealed that the impact of HAP is minimal on the terrestrial network even if the distribution density of HAP increases. But the coverage probability of HAP network gradually decreases as the deployment density increases, because the increase of HAPs will increase the interference to the typical HAP user.  However, the coverage probability of HAP network with directional antenna is less affected by the density of HAPs than the HAP network with omnidirectional antenna. On the one hand, the directional antenna improves the gain of the desired signal. On the other hand, the interference from the side beam is much smaller compared with the desired signal. Thus, the HAPs mounting directional antennas not only improve the coverage probability of HAP network but also enhance the anti-interference ability of HAP network.

\begin{figure}[htb]
	\centering
	\begin{subfigure}[b]{0.45\textwidth}
		\centering
		\includegraphics[scale=0.52]{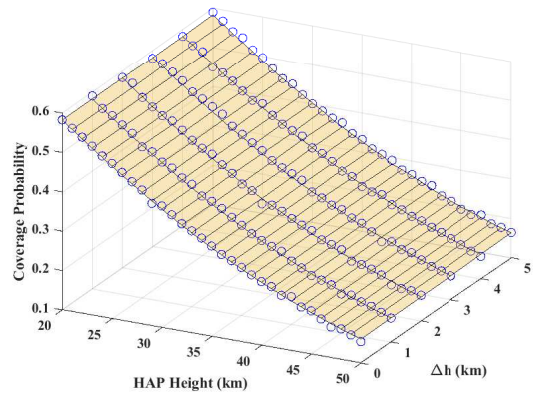}
		\caption{Omnidirectional antenna}
		\label{figure_8_a}
	\end{subfigure}%
	\hfill
	\begin{subfigure}[b]{0.45\textwidth}
		\centering
		\includegraphics[scale=0.52]{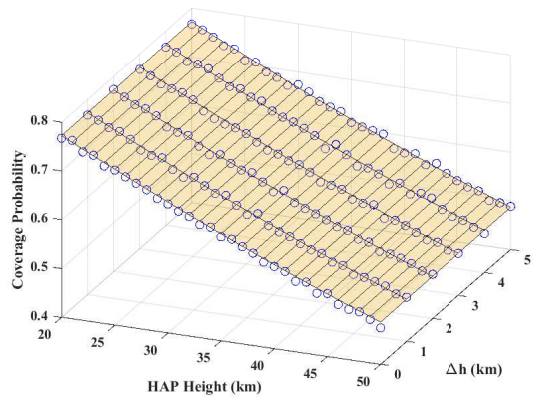}
		\caption{Directional antenna}
		\label{figure_8_b}
	\end{subfigure}%
	\caption{Coverage probability of HAP network vs. HAP height and range of height fluctuation.}
	\label{figure_8}
\end{figure}

Fig. \ref{figure_8} shows the coverage probability of HAP network when the fluctuation height is $1-5$ km, which are compared with the situation without fluctuation. It is discovered that fluctuation in HAP network does not have a large impact on the coverage probability of HAP network due to the negligible range of fluctuation compared to the deployment height. It indicates that the HAP distribution model proposed in this paper has a certain degree of stability and is less affected by the environment.

\begin{figure}[htb]
	\centering
	\begin{subfigure}[b]{0.45\textwidth}
		\centering
		\includegraphics[scale=0.51]{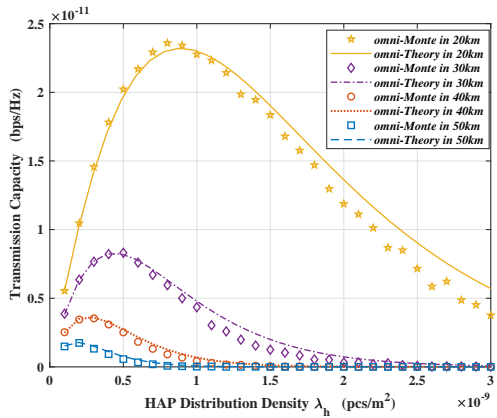}
		\caption{Transmission capacity}
		\label{figure_9_a}
	\end{subfigure}%
	\hfill
	\begin{subfigure}[b]{0.45\textwidth}
		\centering
		\includegraphics[scale=0.51]{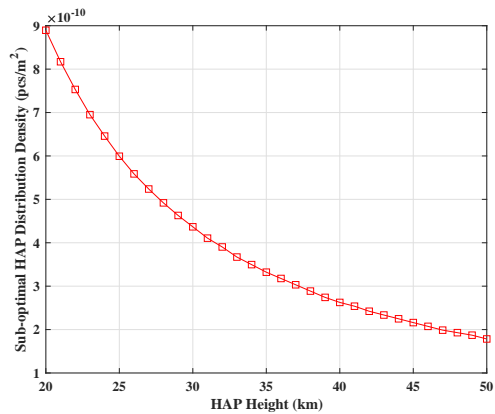}
		\caption{Sub-optimal deployment density}
		\label{figure_9_b}
	\end{subfigure}%
	\caption{Transmission capacity and sub-optimal deployment density under omnidirectional antenna.}
	\label{figure_9}
\end{figure}

Fig. \ref{figure_9} shows the transmission capacity and sub-optimal deployment density of HAP network with omnidirectional antenna. According to Fig. \ref{figure_9_a}, the transmission capacity of HAP network is greatly influenced by the height. In order to obtain a large transmission capacity, HAP should be deployed in the lowest possible height. Meanwhile, there exists an optimal deployment density that maximizes the transmission capacity. The optimization problem \eqref{optimal} can be solved numerically to find the optimal deployment density. 
Fig. \ref{figure_7} reveals that HAP network has less influence on terrestrial network. Thus, the constraint in \eqref{optimal} is very loose, we can search the sub-optimal deployment density of the HAP numerically without considering the constraint, as shown in Fig. \ref{figure_9_b}.

\begin{figure}[htb]
	\centering
	\begin{subfigure}[b]{0.45\textwidth}
		\centering
		\includegraphics[scale=0.51]{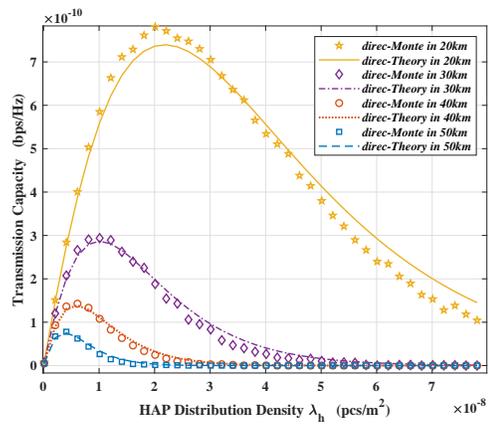}
		\caption{Transmission capacity}
		\label{figure_10_a}
	\end{subfigure}%
	\hfill
	\begin{subfigure}[b]{0.45\textwidth}
		\centering
		\includegraphics[scale=0.51]{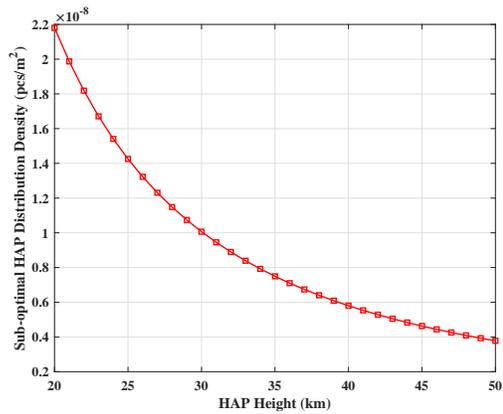}
		\caption{Sub-optimal deployment density}
		\label{figure_10_b}
	\end{subfigure}%
	\caption{Transmission capacity and sub-optimal deployment density under directional antenna.}
	\label{figure_10}
\end{figure}

Fig. \ref{figure_10} shows the transmission capacity and sub-optimal deployment density of HAP network mounted with directional antenna. A comparison between Fig. \ref{figure_10_a} and Fig. \ref{figure_9_a} shows that the transmission capacity of HAP network mounted with directional antenna increases an order of magnitude because the directional antenna improves the coverage probability of HAP network and the anti-interference ability. Fig. \ref{figure_10_b} discovers the sub-optimal deployment density of HAP with the directional antenna at different deployment heights.

\begin{figure}[hbtp]
	\centering
	\includegraphics[scale=0.51]{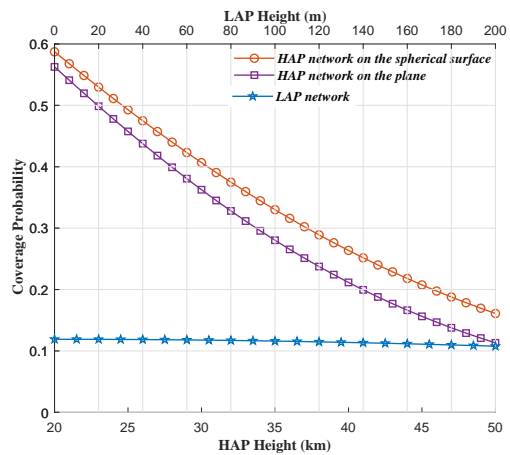}
	\caption{\centering {Comparison between HAP network on the spherical surface, HAP network on the plane, and LAP network. }}
	\label{UAVandHAP}
\end{figure}

Fig. \ref{UAVandHAP} compares the coverage probability of HAP network on the spherical surface with the coverage probability of HAP network on the plane and LAP network with same density. The transmission power of LAP is $5$ W and other parameters come from \cite{PPP3}. For the spherical surface, only HAPs with direct connections to the typical user are considered. However, there is a direct connection between the HAP and its typical user even at an infinite distance for HAP network on the plane. Therefore, the HAP network on the plane will increase interference and reduce the coverage probability. Due to the low height, the communication between LAP and typical user is more likely to be blocked by obstacles, so that its path loss coefficient is large, and Rayleigh fading can be used to model small-scale fading. Generally speaking, the density of HAPs is much smaller than that of LAPs. When the density of LAPs is the same as that of HAPs, the distance between LAP and typical user is relatively large. The received signal in typical user is weak because of the low transmit power. Thus, the coverage probability of LAP network is smaller than that of HAP network.

\section{Conclusion}
This paper  studies the spectrum sharing between HAP network and terrestrial network. To analyze the performance of such a system, the traditional PPP is generalized to curves, surfaces and manifolds firstly. Then, the generalized PPP on surfaces is applied to model the HAPs distributed on the sphere and is analyzed by stochastic geometry and differential geometry. The closed-form expressions for coverage probability of terrestrial network and HAP network with omnidirectional antenna and directional antenna are derived and verified by Monte Carto simulations. The results show that HAP network has less interference to terrestrial network and we can choose a low height and suitable deployment density of HAPs to improve the coverage probability and transmission capacity of HAP network in different height.

\small
\bibliographystyle{ieeetr}

	\vspace{0.5 mm}

\end{document}